\documentstyle [12pt,epsfig] {article}
\def\Journal#1#2#3#4{{#1} {\bf #2}, #3 (#4)}

\def\PLB{{\em Phys. Lett.}  B}
\def\PRT{\em Phys. Rep.}

\def\PRD{{\em Phys. Rev.} D}
\def\NPA{{\em Nucl. Phys.} A}
\def\NPB{{\em Nucl. Phys.} B}
\def\PRC{{\em Phys. Rev.} C}
\def\JPG{{\em J. Phys.} G}

\begin{document}

\begin{titlepage}

\vspace{1cm}

\centerline{\Large \bf Nucleon pole contributions in
$J/\psi\to N \bar{N} \pi$, $p \bar{p} \eta$,}

\centerline{\Large \bf $p \bar{p} \eta^{\prime}$ and $p \bar{p}
\omega$ decays}

\vspace{1cm}

\def\baselinestretch{0.5}

\centerline{\bf {Wei-Hong Liang$^{a,c}$, Peng-Nian Shen$^{d,b,a}$,
Bing-Song Zou$^{b,a,d}$, Amand Faessler$^{e}$ }}

\vspace{0.5cm}

{\small { \flushleft{\bf  $~~~a.$ Institute of High Energy
Physics, Chinese Academy of Sciences,
 P.O.Box 918(4),}
\flushleft{\bf  $~~~~~~$Beijing 100039, China}

\def\baselinestretch{0.5}

\flushleft{\bf  $~~~b.$ China Center of Advanced Science and Technology
 (World Laboratory),}
\flushleft{\bf  $~~~~~~$  P.O.Box 8730, Beijing 100080, China}

\def\baselinestretch{0.5}

\flushleft{\bf  $~~~c.$ Department of Physics, Guangxi Normal University,
Guilin
 541004, China}

\def\baselinestretch{0.5}

\flushleft{\bf $~~~d.$ Center of Theoretical Nuclear Physics, National
Laboratory }

\flushleft{\bf $~~~~~~$of Heavy Ion Accelerator, Lanzhou 730000,
China }

\def\baselinestretch{0.5}

\flushleft{\bf $~~~e.$ Institut f\"{u}r Theoretische Physik,
Universit\"{a}t T\"{u}bingen, }

\flushleft{\bf $~~~~~~$Auf der Morgenstelle 14, D-72076
T\"{u}bingen, Germany }

}}

\vspace{1.5cm}

\def\baselinestretch{1.5}

\centerline{\bf Abstract}

Nucleon pole contributions in $J/\psi \to N \bar N \pi$, $p \bar p
\eta$, $p \bar p \eta^{\prime}$ and $p \bar{p} \omega$ decays are
re-studied. Different contributions due to PS-PS and PS-PV
couplings in the $\pi$-N interaction and the effects of $NN\pi$
form factors are investigated in the $J/\psi \to N \bar N \pi$
decay channel. It is found that when the ratio of $|F_0| /|F_M|$
takes small value, without considering the $NN\pi$ form factor,
the difference between PS-PS and PS-PV couplings are negligible.
However, when the $NN\pi$ form factor is included, this difference
is greatly enlarged. The resultant decay widths are sensitive to
the form factors. As a conclusion, the nucleon-pole contribution
as a background is important in the $J/\psi\to N\bar{N}\pi$ decay
and must be accounted. In the $J/\psi\to N\bar{N}\eta$ and
$N\bar{N}\eta'$ decays, its contribution is less than 0.1$\%$ of
the data. In the $J/\psi\to N\bar{N}\omega$ decay, it provides
rather important contribution without considering form factors.
But the contribution is suppressed greatly when adding the
off-shell form factors. Comparing these results with data would
help us to select a proper form factor for such kind of decay.

\end{titlepage}

\baselineskip 18pt

\section{INTRODUCTION}

Nucleons, as essential building blocks of real world, have been
studied for decades. The members of the nucleon family include
those who are in different excitation modes and even with gluon
contents. The nucleon spectrum investigation would provide us
necessary information for revealing the structure of nucleon
\cite{BK}. So far, in terms of a variety of sources, mostly from
the $\pi N$ elastic and inelastic scattering data, more and more
information on nucleon and its excited states have been
accumulated. However, our knowledge on nucleon family is still far
from completion.

Development of Quantum Chromodynamics (QCD) provides an underlying
theory for the studies of hadrons and their properties. Even so,
nucleon and its family members still cannot strictly be derived
from QCD. The difficulty comes from two sides: the interaction
among quarks and the intrinsic structures of the nucleon and its
family members. To solve this problem in a more efficient way,
various QCD inspired models have been proposed. As a result, many
nucleon resonances ($N^*$) have been predicted.

On the experimental side, searching $N^*$s has been an very
important project in past years. The results were mainly extracted
from the $\pi N$ scattering data. Up to now, many nucleon
resonances have been found. Yet, still some $N^*$ states which
were predicted by widely accepted nucleon models, such as quark
models \cite{Isgurk}, have not been seen in the $\pi N$ channel.
Whether these so-called ``missing resonances" couple weakly to the
$\pi N$ channel \cite{Faiman, Capstick1}, so that we should
propose other means to search them? Or, if the quark model
predicts too many resonances so that the model itself should
further be modified? Or, there may exist the hybrid structure or
the di-quark structure? All these puzzles motivate intensive
investigations in both experimental side and the theoretical side.

\begin{figure}[htbp]
\begin{center}
\epsfxsize=3.0in
\epsfbox{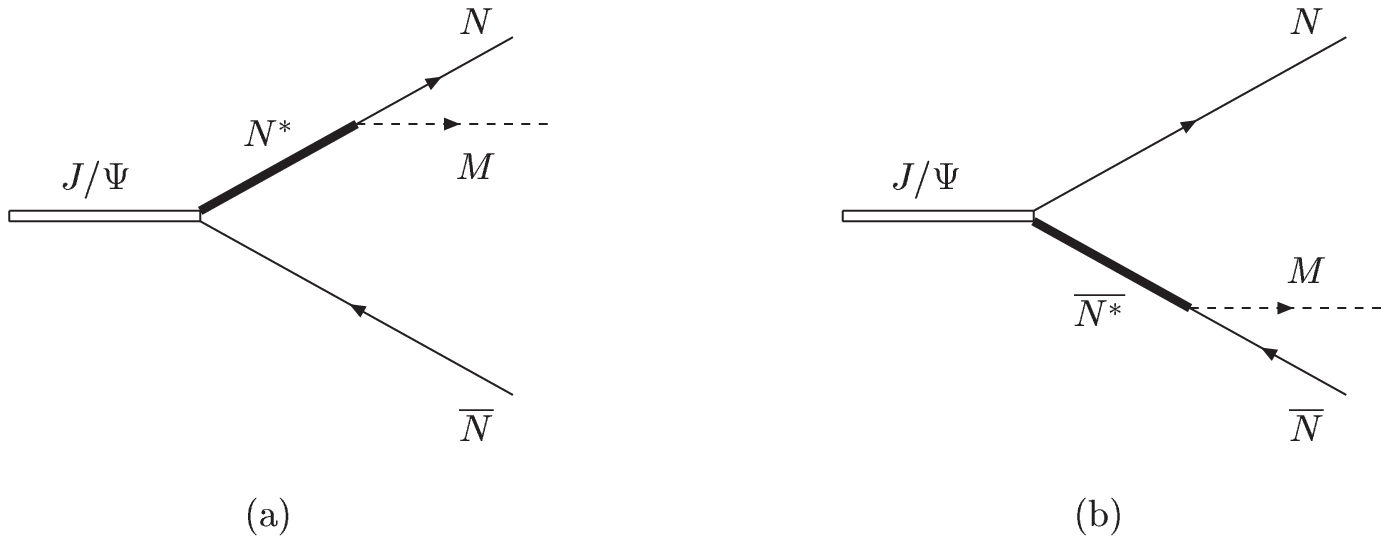}
\caption{\small {$N^*$-pole diagrams for $J/\psi\to M N \bar{N}$
decay.}}
\label{Nstar}
\end{center}
\end{figure}

\begin{figure}[htbp]
\begin{center}
\epsfxsize=3.0in
\epsfbox{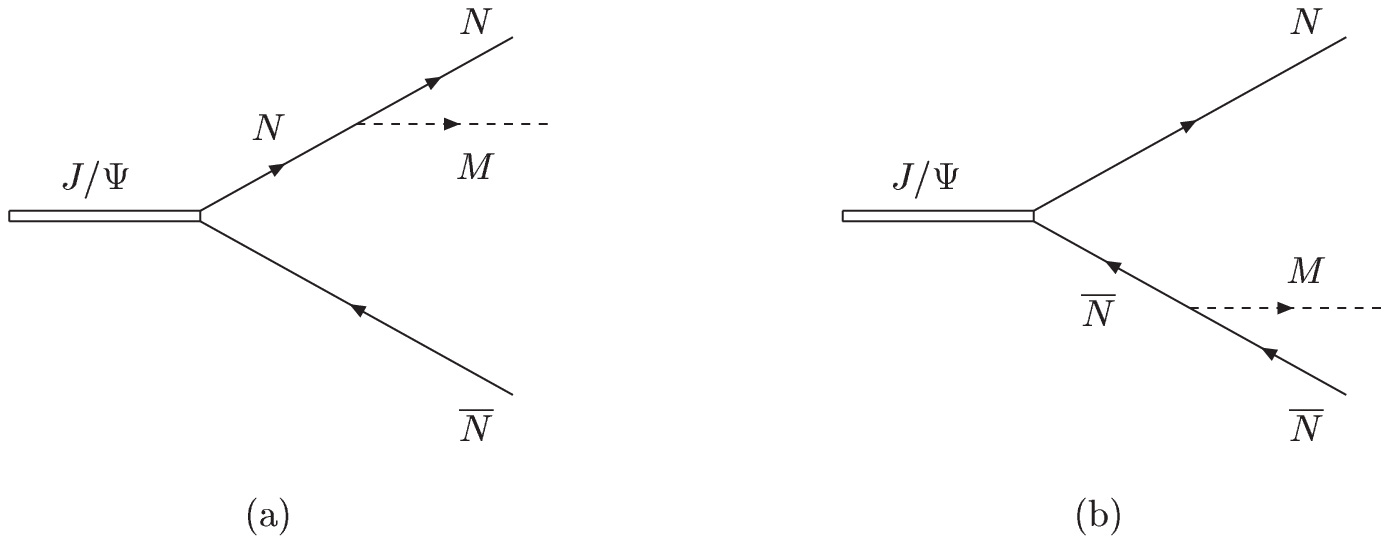}
\caption{\small {Nucleon-pole
diagrams for $J/\psi \to M N \bar{N}$ decay.}}
\label{Mpole}
\end{center}
\end{figure}

In recent years, a large number of experiments on $N^*$ physics
have been carried out at new facilities such as CEBAF at JLAB,
ELSA at Bonn, GRAAL at Grenoble and SPring-8 at JASRI. Now, 58
million $J/\psi$ events have been collected at Beijing
Electron-Positron Collider (BEPC).  The two-step decay process
$J/\psi \to N^* \bar{N}\to M N \bar{N}$, where $M$
refers to meson, can be another excellent source for studying
light baryon resonances with many advantages \cite{Zoubs1,
Zoubs2}. Corresponding Feynman diagrams are shown in Fig.
\ref{Nstar}.
%
%%%%%%%%%%%%%%% Fig. 1 %%%%%%%%%%%%%%%%%%%%%%%%%%%%%%%%%%%%%%
%%%%%%%%%%%%%%%%
%
It should be mentioned that the nucleon-pole diagrams (shown in
Fig. \ref{Mpole}) would also contribute as a background in the
$N^*$ study via $J/\psi\to M N \bar{N}$ decays. For light
mesons, especially for pion, nucleon-pole contributions might be
sizable and should not be ignored.
%
%%%%%%%%%%%%%%%%%%%%% Fig. 2 %%%%%%%%%%%%%%%%%%%%%%%%%%%

In order to extract a more accurate and reliable conclusion from
the $J/\psi$ hadronic decay data, it is necessary to study the
nucleon-pole contributions in those decay channels. By analyzing
$J/\psi \to  p \bar p \pi^0$ data, R.Sinha and S.Okubo
\cite{Sinha} pointed out that in the $J/\psi \to  p \bar p \pi^0$
decay, the p-pole contribution dominates in the soft pion limit,
and the $N^*$-pole contribution becomes important at the large
pion energy region. In the $J/\psi \to  p \bar p \eta$ and $p
\bar{p} \eta^{\prime}$ decays, if one considers the p-pole
contribution only, the extracted $g_{\eta N \bar{N}}/g_{\pi N
\bar{N}}$ value would be much smaller than that from the
experimental decay widths of the $J/\psi \to  p \bar p \pi^0$, $p
\bar p \eta$ and $p \bar{p} \eta^{\prime}$ processes. Due to the
fact that the decay rates of $\Gamma (N^*\to \eta N)$ are rather
large for both $N^*(1440)$ and $N^*(1535)$, the $N^*$-pole
contribution must governing $J/\psi \to  p \bar p \eta$ and $p
\bar{p} \eta^{\prime}$ decays. In the $J/\psi\to p\bar{p}\omega$
decay, the p-pole contribution only gives 1/10 of the experimental
decay rate. Therefore, in order to obtain a reliable information
about $N^*$ via $J/\psi \to  p \bar p M$ decays, one should
carefully consider the p-pole contribution as the part of the
background. In this work, we calculate the nucleon-pole
contributions in the $J/\psi \to  N \bar N \pi$, $p \bar p \eta$,
$p \bar p \eta^{\prime}$ and $p\bar{p}\omega$ decays with various
hadronic form factors. In general, the main propose of this paper
is to emphasize the importance of the contribution of the
nucleon-pole diagram in studying $N^*$'s via various
$J/\psi\rightarrow N\bar{N} M $ processes and to provide a
direction that how big the deviation would be when the vertex form
factors are applied in data analysis.

The paper is organized in the following way: in the next section,
the nucleon-pole contributions in the $J/\psi\to N\bar{N}\pi$,
$p\bar{p}\eta$ and $p\bar{p}\eta^{\prime}$ decays with and without
form factors is systematically studied. The nucleon-pole
contribution in the $J/\psi\to N\bar{N}\omega$ decay is
demonstrated in section 3, and in section 4, the conclusion is
given.

\section{Nucleon pole contributions in  $J/\psi \to  N \bar N \pi$, $p
\bar p \eta$ and $p \bar p \eta^{\prime}$ decays}

Firstly, we take the $J/\psi\to N \bar{N}\pi$ channel as a
sample to analyze cautiously the off-shell effect through
different $NN\pi$ couplings and various form factors. And then, we
discuss the results in the $J/\psi\to N \bar{N}\eta$ and
$N \bar{N}\eta^\prime$ channels.

\subsection{
Nucleon pole contributions by using different $NN\pi$ couplings in the
$J/\psi\to N \bar{N}\pi$ decay}

%
%%%%%%%%%%%%%%%%%%%%%  Fig. 3 %%%%%%%%%%%%%%%%%%%%%%
\begin{figure}[htbp]
\begin{center}
\epsfxsize=3.0in
\epsfbox{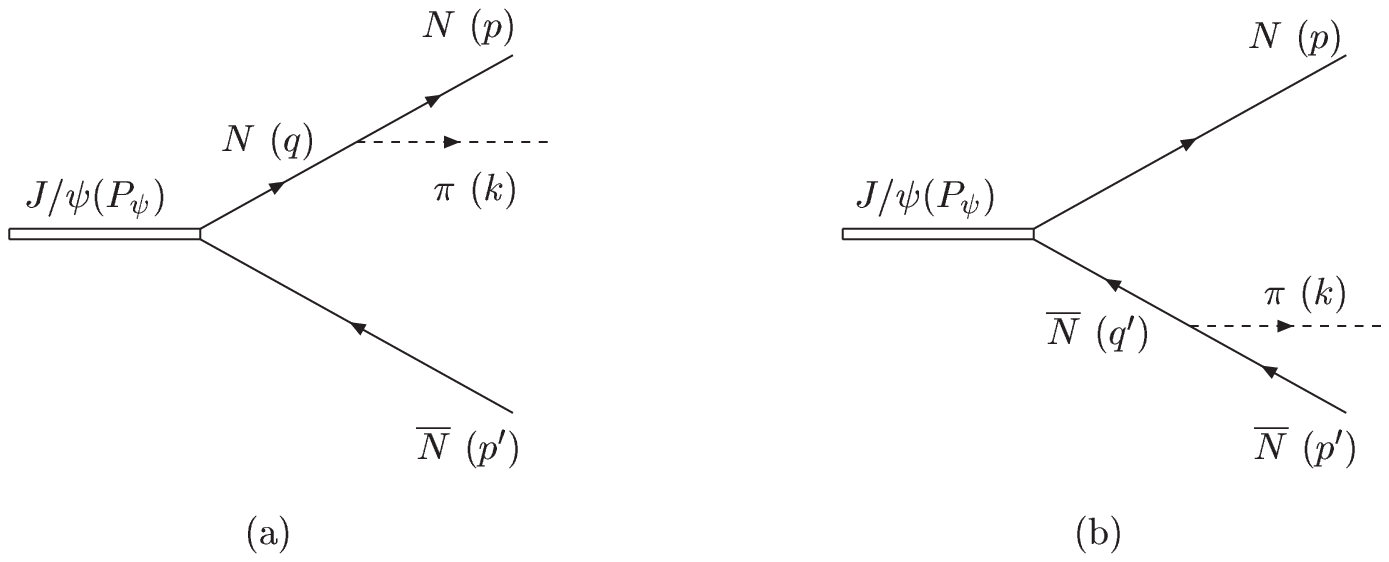}
\caption{\small
{Nucleon-pole diagrams for $J/\psi \to \pi N \bar{N}$ decay.}}
\label{pi_N}
\end{center}
\end{figure}

The nucleon-pole diagrams for $J/\psi \to \pi N \bar N$
are shown in Fig. \ref{pi_N}, with $q = p+k =P_{\psi}-p^{\prime}$
and $q^{\prime} = p^{\prime}+k = P_{\psi}-p$. In the case of very
low energy of pion, the dominant contribution to the
$J/\psi\to N \bar{N}\pi$ decay comes from the nucleon-pole
diagram. However, when the energy of pion becomes large, the
contribution of the nucleon-pole diagram is evidently larger than
the data. Thus, the off-shell effect of the nucleon propagator
should be carefully studied. Generally, the $J/\psi\to N
\bar{N}$ interaction can be written as
\begin{equation}
H_{\psi} = \bar N ~[F_M \gamma^{\mu} + \frac{1}{2m}
F_0 {(p-p^{\prime})}^{\mu} ] ~N~ \epsilon_{\mu}(P_{\psi}),
\end{equation}
where m is the mass of the nucleon, $P_{\psi}$, $p$ and
$p^{\prime}$ are the four momenta of $J/\psi$, $N$ and $\bar N$,
respectively, and $\epsilon_{\mu}(P_{\psi})$ denotes the
polarization vector of $J/\psi$.  Dimensionless real decay
constants $F_M$ and $F_0$ can be determined by the experimental
data of the two-body decay $J/\psi \to p \bar p$. There
are two forms for pion-nucleon interaction which are widely
employed in literatures. One is in pseudoscalar-pseudoscalar
(PS-PS) form:
\begin{equation}
H_1=ig_{N \bar{N}\pi} \bar{N} \gamma_5  \vec{\tau} N \vec{\pi},
\label{h1}
\end{equation}
and the other is in
pseudoscalar-pseudovector (PS-PV) form:
\begin{equation}
H^{\prime}_1=\frac{1}{2m} g_{N \bar{N}\pi} \bar{N} \gamma_5 \gamma_{\mu}
\vec{\tau} N \partial^{\mu} \vec{\pi},
\label{h1prime}
\end{equation}
where $\vec{\tau}$ is the isospin Pauli matrix, and $g_{N\bar{N} \pi}$ is the
pion-nucleon coupling constant with \cite{Sinha}
\begin{equation}
{(g_{N \bar{N}\pi})}^2/4\pi \simeq 14.8 ~.
\end{equation}

When the intermediate nucleon is on-shell, the decay amplitude of
Fig. \ref{pi_N} in the PS-PS coupling
 $\pi$-N  interaction can be derived as
\begin{eqnarray}
{\cal{M}}^{on}_{PS} &=& i g_{N\bar{N} \pi}  \bar{u}(p) \gamma_5~
\left[~ F_M \left( \frac{\slash{\hskip -2.0mm}k ~\slash{\hskip -2.0mm}
\epsilon}{2p\cdot k +k^2}
-\frac{\slash{\hskip -2.0mm} \epsilon
~\slash{\hskip -2.0mm}k}{2p^{\prime}\cdot k +k^2}
\right) \right.  \nonumber \\
&~& \left. +\frac{F_0}{m} ~\slash{\hskip -2.0mm}k \left(
\frac{p \cdot \epsilon}{2p^{\prime}\cdot k +k^2}
-\frac{p^{\prime} \cdot \epsilon}{2p\cdot k +k^2}
\right)
\right]~v(p^{\prime})  \label{ps-on} \\
& \equiv & {\cal{M}}_{PS}. \nonumber
\end{eqnarray}
It also can easily be proved that the decay amplitude in the PS-PV
coupling case takes the same form, namely
\begin{equation}
{\cal{M}}^{on}_{PV}={\cal{M}}_{PS}.
\label{pv-on}
\end{equation}
Thus, no matter PS-PS coupling or PS-PV coupling is employed, the
yielded decay amplitudes would be exactly the same.

\begin{figure}[htbp]
\begin{center}
\includegraphics[width=12cm]{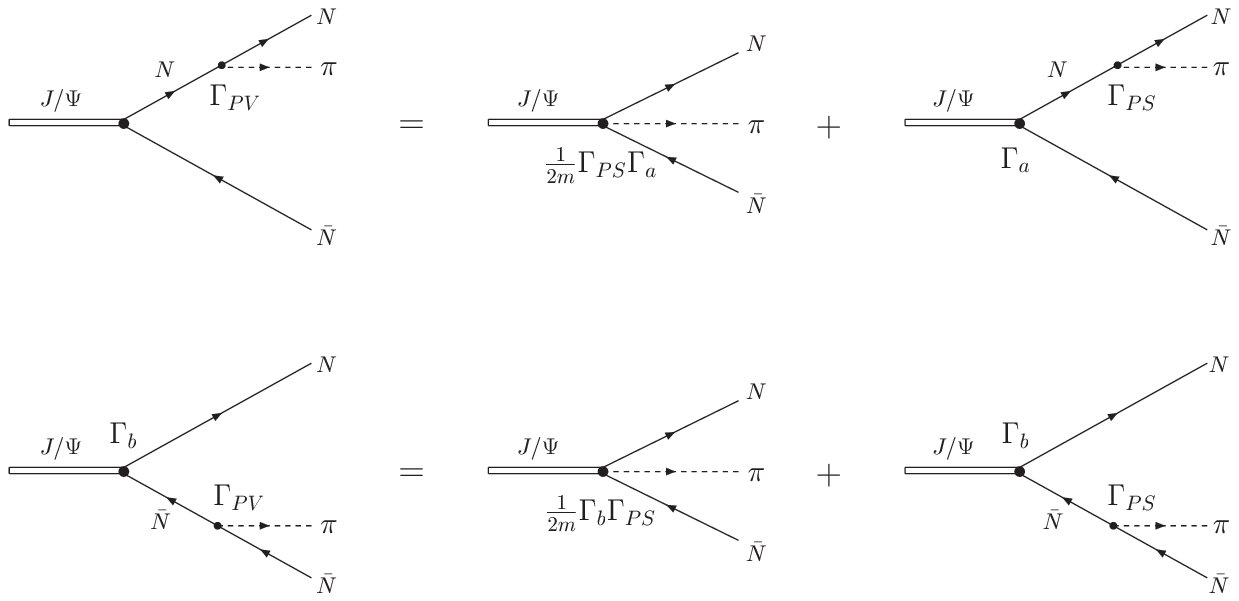}
\caption{}
\label{fig:plus-fig}
\end{center}
\end{figure}

It can further be verified that when the intermediate nucleon is
off-shell, the decay amplitude of Fig. \ref{pi_N} in the PS-PS
coupling case still takes the same form as in the on-shell case
\begin{equation}
{\cal{M}}^{off}_{PS}={\cal{M}}_{PS},
\label{ps-off}
\end{equation}
but in the PS-PV coupling case, it has additional terms:
\begin{equation}
{\cal{M}}^{off}_{PV,a} = \frac{ig_{N\bar{N} \pi}}{2m} \bar{u}(p)
\gamma_5 [F_M \gamma^{\mu} \epsilon_{\mu}+\frac{1}{2m}F_0
{(q-p^{\prime})}^{\mu}
\epsilon_{\mu}]v(p^{\prime})+{\cal{M}}_{PS,a} \label{pv-a},
\end{equation}
\begin{equation}
{\cal{M}}^{off}_{PV,b} = \frac{ig_{N\bar{N} \pi}}{2m} \bar{u}(p)
[F_M \gamma^{\mu} \epsilon_{\mu}+\frac{1}{2m}F_0
{(p-q^{\prime})}^{\mu} \epsilon_{\mu}]\gamma_5
v(p^{\prime})+{\cal{M}}_{PS,b}~, \label{pv-b}
\end{equation}
and
\begin{equation}
{\cal{M}}_{PS,a} + {\cal{M}}_{PS,b} = {\cal{M}}_{PS}~,
\label{pv-ab}
\end{equation}
where the subscripts $a$ and $b$ denote the decay amplitudes of
Fig. \ref{pi_N}(a) and Fig. \ref{pi_N}(b), respectively.
Eqs.(\ref{pv-a}) and (\ref{pv-b}) can also be expressed
diagrammatically as Fig. \ref{fig:plus-fig},
%
%%%%%%%%%%%%%%%%%%%%% Fig. 4 %%%%%%%%%%%%%%%%%%%%%%%
where the vertices $\Gamma_a$, $\Gamma_b$, $\Gamma_{PS}$ and
$\Gamma_{PV}$ are $\Gamma_a = F_M \gamma^{\mu} + \frac{1}{2m} F_0
{(q-p^{\prime})}^{\mu}$, $\Gamma_b = F_M \gamma^{\mu} +
\frac{1}{2m} F_0 {(p-q^{\prime})}^{\mu}$, $\Gamma_{PS} = i g_{N
\bar{N} \pi} \gamma_5$ and $\Gamma_{PV} = \frac{i}{2m} g_{N
\bar{N} \pi} \gamma_5 \gamma_{\mu} k^{\mu}$, respectively. The
total decay amplitude for Fig. \ref{pi_N} is then obtained by
summing over Eq.(\ref{pv-a})  and Eq.(\ref{pv-b})
\begin{eqnarray}
{\cal{M}}^{off}_{PV} &=& {\cal{M}}^{off}_{PV,a} +{\cal{M}}^{off}_{PV,b}
\nonumber \\
&=&\frac{ig_{N\bar{N} \pi}}{2m^2}F_0 \bar{u}(p) {(p-p^{\prime})}^{\mu}
\epsilon_{\mu}
\gamma_5 v(p^{\prime}) +{\cal{M}}_{PS}  \label{off-pv}\\
&\equiv& {\cal{M}}_{PV}. \nonumber
\end{eqnarray}

It is clear that in the $J/\psi\to N \bar{N}\pi$ process,
when $\pi$-N interaction takes the PS-PV coupling form, the decay
amplitude would receive not only the same contribution from PS-PS coupling, but
also an extra contribution from contact term. Moreover, the difference of the
decay amplitudes in the PS-PS and PS-PV coupling cases only relate to
$|F_0|$, and are more distinct at the large value of $|F_0|$.

The differential decay rate can be formulated by summing over
possible spin states of final nucleon and anti-nucleon
\begin{eqnarray}
d\Gamma_{PS}(J/\psi \to N \bar{N} \pi)&=&
\frac{{2\pi}^4}{2M_{\psi}}{|{\cal{M}}_{PS}|}^2  d\Phi_3 (P_{\psi}; p,
p^{\prime}, k) \nonumber \\
&=& {(2\pi)}^4 \frac{2g^2_{N \bar{N}
\pi}}{M_{\psi}}~[~{|F_M|}^2 A_{PS,1}+{|F_0|}^2 A_{PS,2} \nonumber \\
&~&+Re(F^{*}_0 F_M)A_{PS,3}~]~ d\Phi_3 (P_{\psi}; p, p^{\prime}, k) ,
\label{appen1}
\end{eqnarray}
\begin{eqnarray}
d\Gamma_{PV}(J/\psi \to N \bar{N} \pi)&=&
\frac{{2\pi}^4}{2M_{\psi}}{|{\cal{M}}_{PV}|}^2  d\Phi_3 (P_{\psi}; p,
p^{\prime}, k) \nonumber \\
&=& {(2\pi)}^4 \frac{2g^2_{N \bar{N}
\pi}}{M_{\psi}}~[~{|F_M|}^2 A_{PS,1}+{|F_0|}^2
(A_{PS,2}+A_{PV,2}) \nonumber \\
&~&+Re(F^{*}_0 F_M)(A_{PS,3}+A_{PV,3})~]~
d\Phi_3 (P_{\psi}; p, p^{\prime}, k),
\label{appen2}
\end{eqnarray}
with $M_{\psi}$ being the mass of $J/\psi$ and
\begin{equation}
d\Phi_3 (P_{\psi}; p, p^{\prime}, k) = \delta^4 (P_{\psi}-p-p^{\prime}-k)
\frac{d^3 p}{{(2 \pi)}^3 2p_0} \frac{d^3 p^{\prime}}{{(2 \pi)}^3
2p^{\prime}_0} \frac{d^3 k}{{(2 \pi)}^3 2k_0}
\end{equation}
being an element of three-body phase space. The explicit
expressions for $A_{PS, i}(i=1,2,3)$ and $A_{PV, i}(i=2,3)$ are
shown in the Appendix. Again, it is found from Eqs.(\ref{appen1})
and (\ref{appen2}) that the ${|F_M|}^2$ dependent term does not
contribute to the difference between $d\Gamma_{PV}$ and
$d\Gamma_{PS}$. If $|F_0| =0$, the differential decay rates in the
PS-PS and PS-PV coupling cases are absolutely identical.

The value of $|F_0|/|F_M|$ can be determined in the following way
\cite{Sinha}. In the realistic calculation, one usually adopts the
electric coupling parameter $F_E$ and the magnetic coupling
parameter $F_M$ instead of  $F_0$ and $F_M$ used above. $F_0$
can be expressed by
\begin{equation}
F_0 = \frac{4m^2}{M_{\psi}-4m^2} (F_M -F_E).
\label{fe}
\end{equation}
Then the squared amplitude for $J/\psi \to p \bar p$ decay
can be written as
\begin{equation} \overline{{|{\cal M}|}^2}=C_0
~(1 + \alpha cos^2\theta),
\end{equation}
with
\begin{equation}
C_0=m^2{|F_M|}^2+4m^2{|F_E|}^2 , ~~~
 \alpha =
[{|F_M|}^2-\frac{4m^2}{M^2_{\psi}}{|F_E|}^2] ~/~ [{|F_M|}^2
+\frac{4m^2}{M^2_{\psi}}{|F_E|}^2] . \label{alpha}
\end{equation}
By measuring the angular distribution of the $J/\psi \to p
\bar p$ decay, one obtains $\alpha=0.62 \pm 0.11$ \cite{DM2}.
Consequently, $|F_E|/|F_M|=0.80 \pm 0.14$. Assuming
\begin{equation}
\frac{F_E}{F_M} = \frac{|F_E|}{|F_M|} ~e^{i\delta}~,
\label{eq:phasefactor}
\end{equation}
one can easily extract the value of $|F_0|/|F_M|$ by
\begin{equation}
\frac{F_0}{F_M} = \frac{4m^2}{M^2_{\psi}-4m^2} \left[ 1-
\frac{|F_E|}{|F_M|}~e^{i\delta} \right]~.
\end{equation}
Taking $\delta = 0$, $\frac{\pi}{2}$ and $\pi$, we have
\begin{equation}
\frac{|F_0|}{|F_M|}  =\left\{
\begin{array}{ll}
0.12 \pm 0.08  &~ ~for ~~\delta~=0,\\
0.74 \pm 0.08  & ~~for ~~\delta~= {\textstyle\frac{\pi}2},\\
1.04 \pm 0.08  & ~~for ~~\delta~=\pi.\\
\end{array}
\right. \label{eq:F0FM}
\end{equation}

The effect of the ratio $|F_0|/|F_M|$ on $\Gamma_{PS}$ and
$\Gamma_{PV}$ can be exhibited  by taking $|F_0|/|F_M| = 0, 0.12, 0.74,
1.04$ in our calculation. The branching ratio (BR) of
the decay widths for Fig. \ref{pi_N} in the PS-PS and PS-PV cases
are{\footnote {The corresponding ratios in Ref. \cite{Sinha} are
larger than ours due to their large deviation in the phase space
integration. }}
\begin{equation}
\frac{\Gamma_{PS}(J/\psi \to p \bar{p}\pi^0)}{\Gamma(J/\psi \to p
\bar{p})}  =\left\{
\begin{array}{ll}
0.556  &~ ~for ~|F_0|/|F_M|~=0,\\
0.561  & ~~for ~|F_0|/|F_M|~=0.12,\\
0.688    & ~~for ~|F_0|/|F_M|~=0.74,\\
0.815  & ~~for ~|F_0|/|F_M|~=1.04,\\
\end{array}
\right.
\label{R1}
\end{equation}
and
\begin{equation}
\frac{\Gamma_{PV}(J/\psi \to p \bar{p}\pi^0)}{\Gamma(J/\psi \to p
\bar{p})}  =\left\{
\begin{array}{ll}
0.556  &~ ~for ~|F_0|/|F_M|~=0,\\
0.529  &~ ~for ~|F_0|/|F_M|~=0.12,\\
0.475  & ~~for ~|F_0|/|F_M|~=0.74,\\
0.421  & ~~for ~|F_0|/|F_M|~=1.04,\\
\end{array}
\right.
\label{R2}
\end{equation}
respectively. Comparing with the empirical ratio \cite{PDG2000}
\begin{equation}
\frac{\Gamma(J/\psi \to p \bar{p}\pi^0)} {\Gamma(J/\psi
\to p \bar{p})} = 0.51 \pm 0.04,
\label{exp}
\end{equation}
one sees that the resultant BRs in Eqs.(\ref{R1}) and (\ref{R2})
are very close to the data. It indicates that the BR of the
$J/\psi \to N \bar N {\pi}$ decay is dominated by the
nucleon-pole diagrams of Fig. \ref{pi_N} without including the
hadronic form factor.
Of course, the $N^*$-pole will also
contribute. However, if one would use the data of $J/\psi
\to N \bar N \pi$ decay to study $N^*$, one cannot get
meaningful information until the nucleon-pole contribution is
considered.

One can also find that the difference due to
different $\pi$-N couplings becomes larger when $|F_0|/|F_M|$
ratio increases. This is because that in the PS-PS coupling case,
both the $F_M$-dependent term and the $F_0$-dependent term
contribute positively, but in the PS-PV coupling case, the
$F_M$-dependent term keeps the same contribution and the
$F_0$-dependent term gives a negative contribution. Therefore,
large $|F_0|/|F_M|$ value would make the difference larger. For
instance, with $|F_0|/|F_M|=1.04$, the ratio of $\Gamma_{PS} (J/\psi
\to p \bar p \pi^0) / \Gamma (J/\psi \to p \bar p)$ is
almost twice of $\Gamma_{PV} (J/\psi \to p \bar p \pi^0) /
\Gamma (J/\psi \to p \bar p)$.

\iffalse
With $|F_0|/|F_M| = 0.12$, the total decay rates $\Gamma (J/\psi
\to N \bar{N} \pi)$ for Fig. \ref{pi_N} are
\begin{equation}
\frac{\Gamma_{PS} (J/\psi \to p \bar p \pi^0)}{\Gamma
(J/\psi \to p \bar p )} = 0.563~ ,
\label{eq:piBRPS2}
\end{equation}
%
\begin{equation}
\frac{\Gamma_{PV} (J/\psi \to p \bar p \pi^0)}{\Gamma
(J/\psi \to p \bar p )} = 0.529~ ,
\label{eq:piBRPV2}
\end{equation}
%
%
and
\begin{equation}
\frac{\Gamma_{PV} (J/\psi \to p \bar p \pi^0)}{\Gamma_{PS} (J/\psi
\to p \bar p \pi^0)} \simeq 0.940~. \label{eq:PV:PS-NoFF}
\end{equation}
This shows that when $|F_0|/|F_M|=0.12$, the results by using
either PS-PS or PS-PV couplings are comparable. We can conclude
that the nucleon-pole diagram dominates the $J/\psi \to N
\bar N \pi$ decay. Of course, the $N^*$-pole will also
contribute. However, if one would use the data of $J/\psi
\to N \bar N \pi$ decay to study $N^*$, one cannot get
meaningful information until the nucleon-pole contribution is
considered.
\fi

\vspace{0.5cm}

\subsection{Off-shell effect with various form factor in the
$J/\psi \to N \bar N \pi$ decay }

Normally, a hadronic form factor is applied to the
meson-baryon-baryon ($MBB^{\prime}$) vertices because of the inner
quark-gluon structure of hadrons. It is well known that form
factor plays an important role in many physics processes, for
example, the NN interaction models \cite{Machleidt1}, NN
scattering \cite{Gross92}, $\pi N$ scattering \cite{Gross93,
Pearce91, Schutz94}, pion photoproduction \cite{Sato96}, vector
meson photoproduction \cite{Q.Zhao} and etc.. However, due to the
difficulties in dealing with nonperturbative QCD (NPQCD) effects,
the form factors are commonly adopted phenomenologically.

The most commonly used form factors for meson-nucleon-nucleon
vertices are monopole form factor and dipole form factor
\cite{Liu}:
\begin{equation}
F_1(q^2)=\frac{\Lambda^2+m^2}{\Lambda^2+q^2},
\label{F1}
\end{equation}
\begin{equation}
F_2(q^2)=\frac{\Lambda^4+m^4}{\Lambda^4+q^4},
\label{F2}
\end{equation}
where m and q are the mass and the four-momentum of the
intermediate particle, respectively, and $\Lambda$ is the
so-called cut-off momentum that can be determined by fitting the
experimental data. The monopole form factor is mainly used in the
$\pi$-N and N-N interactions, while the dipole one is usually
applied to the N-N interaction. The values of $\Lambda$ is
different process by process. A typical value of $\Lambda$ for a
monopole form factor in the Bonn potential is in the region of 1.3
$\sim$ 2~ GeV \cite{Machleidt1}, and for $\pi$-N interaction is
about 1.35~GeV. Frankfurt and Strikman \cite{Frankfurt} analyzed
the deep inelastic scattering (DIS) of leptons from nucleons and
showed that the DIS data support a $\pi NN$ monopole form factor
with $\Lambda \leq 650 MeV$.

The exponential form factor is also a frequently used
meson-nucleon-nucleon form factor \cite{exp},
\begin{equation}
F_3(q^2)=e^{-|q^2-m^2|/\Lambda^2}.
\label{F3}
\end{equation}
Form factor can also take the following form \cite{new, Lee1}:
\begin{equation}
F_4(q^2)=\frac{1}{1+{(q^2-m^2)}^2/\Lambda^4}.
\label{F4}
\end{equation}
Moreover, in the study of the photoproduction of meson, T.-S. H.
Lee et al. \cite{Lee1, Lee2} chose a form factor with the form of
\begin{equation}
F_5({\bf q}^2)= exp[-({\bf q}^2 - {\bf q}^2_0) / \Lambda^2],
\label{F5}
\end{equation}
where ${\bf q}$ and ${\bf q}_0$ are the three-momentum vectors of
the intermediate nucleon at the energy of $\sqrt{s} (s=q^2)$ and
at the nucleon pole position, respectively.

It should be mentioned that all the form factors mentioned above
are normalized to unity when the intermediate nucleon is on its
mass shell.

\begin{figure}[htbp]
\vspace{0.6cm}
\begin{center}
\includegraphics[width=7.2cm]{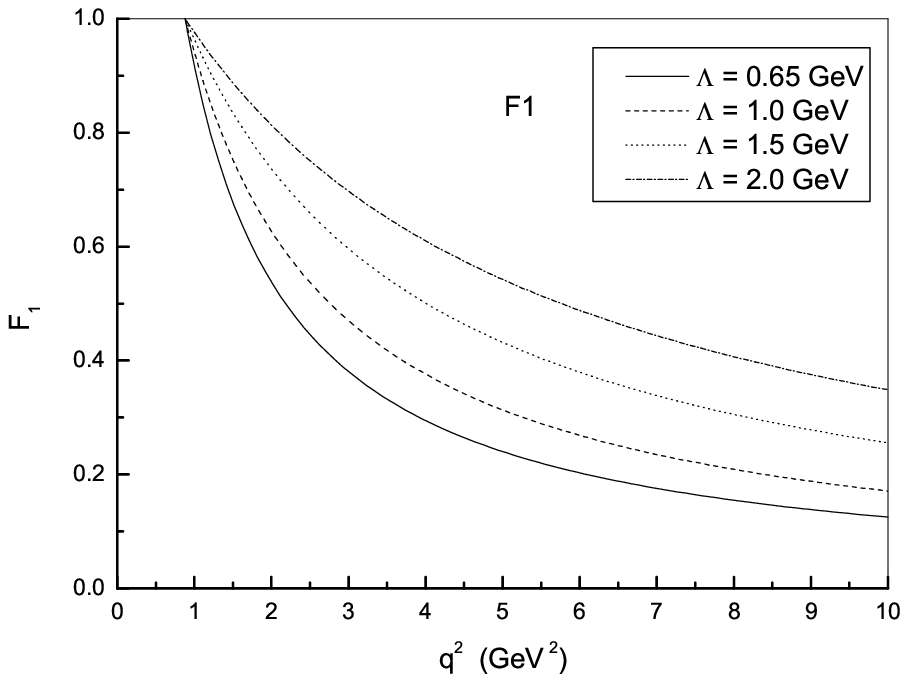}
\includegraphics[width=7.2cm]{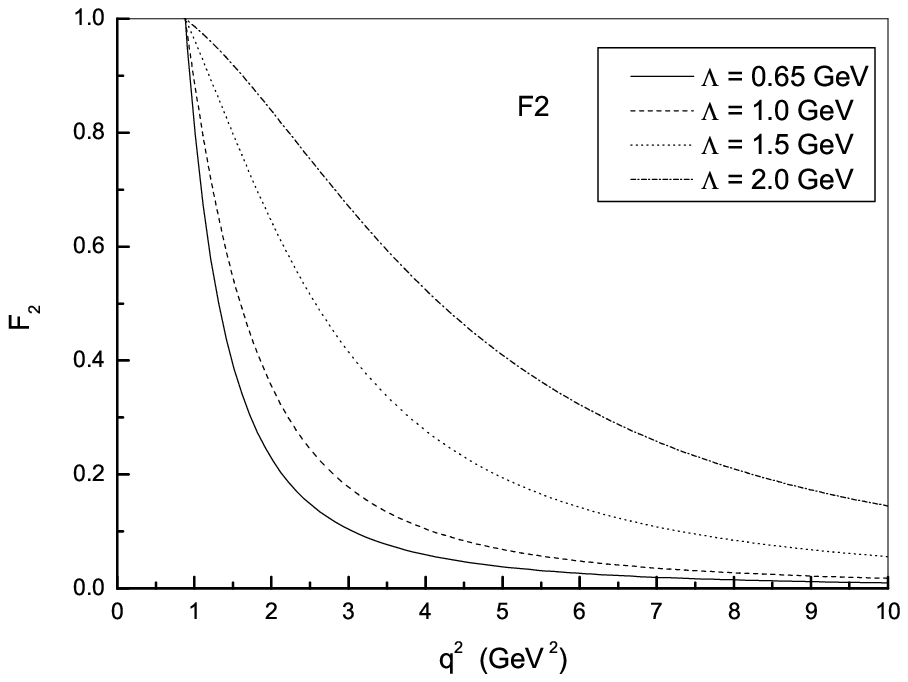}
\includegraphics[width=7.2cm]{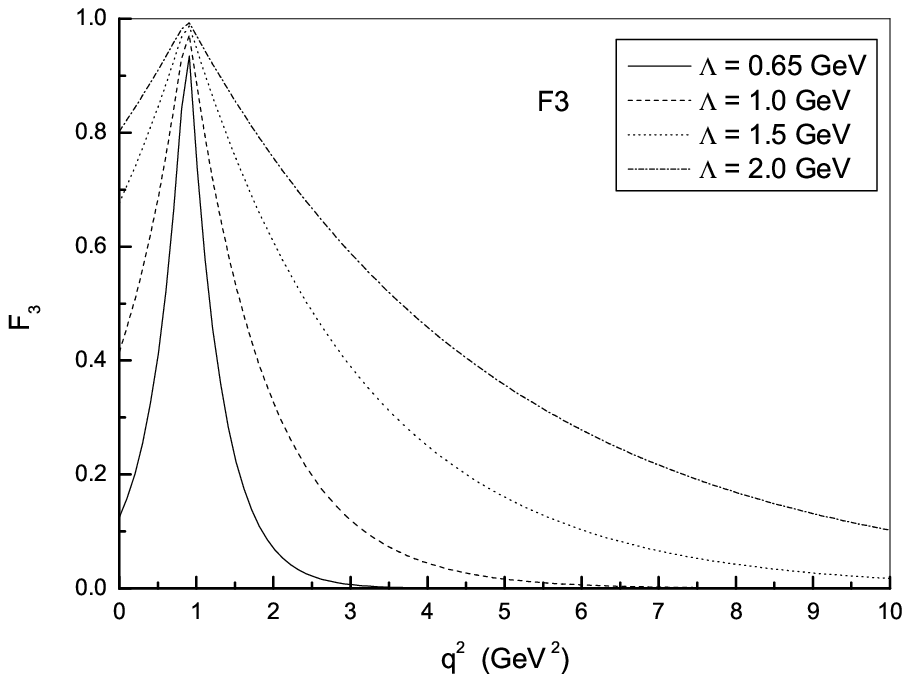}
\includegraphics[width=7.2cm]{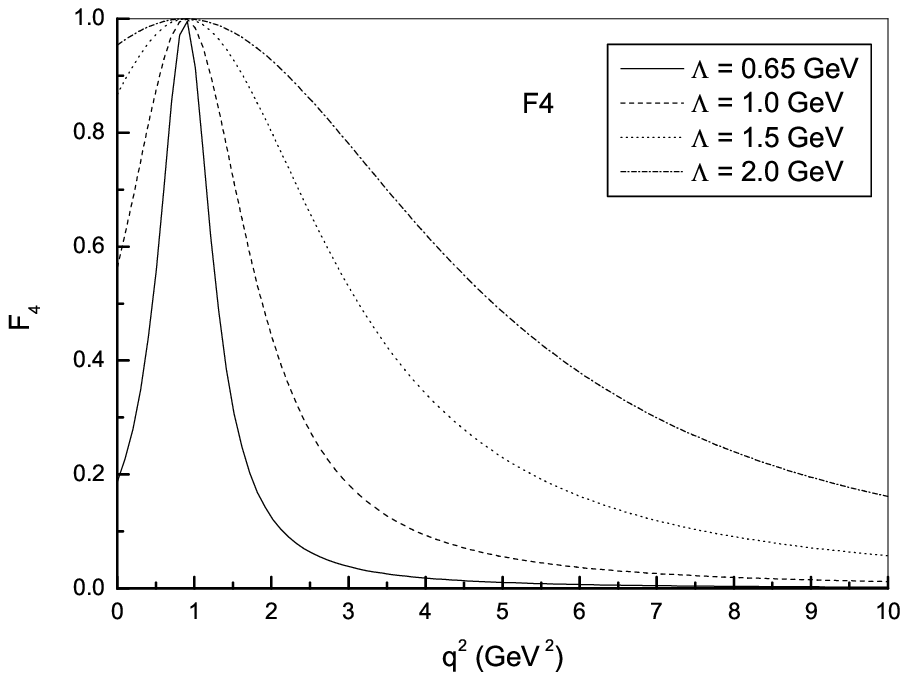}
\includegraphics[width=7.2cm]{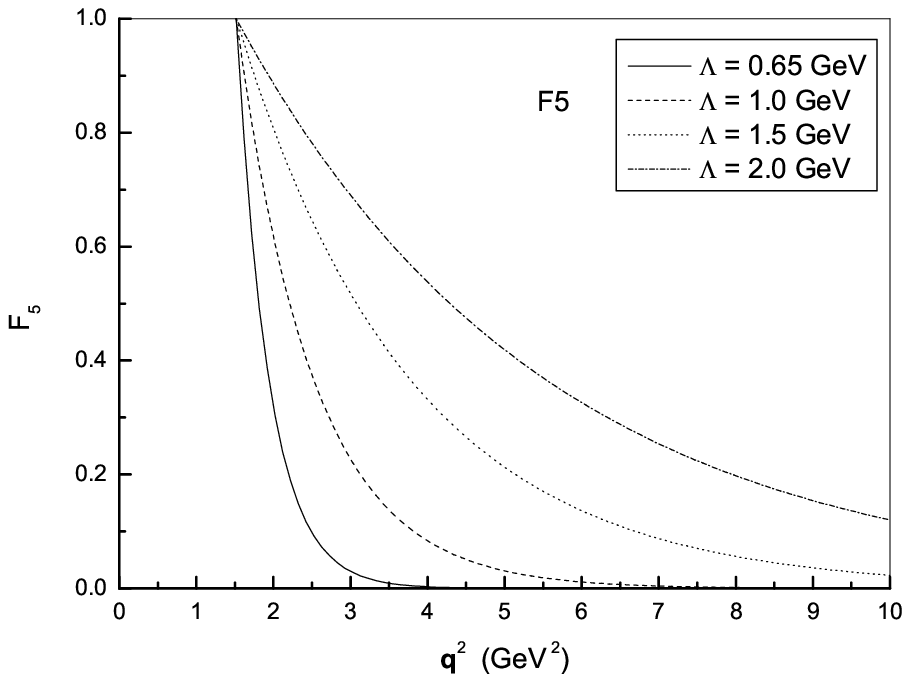}
\includegraphics[width=7.2cm]{empty.eps}
\caption{\small {The momentum-dependence
of form factors $F_1 \sim F_5$ with different $\Lambda$ values.}}
\label{fig:FormFactor-FF}
\end{center}
\end{figure}

\begin{figure}[htbp]
\vspace{0.6cm}
\begin{center}
\includegraphics[width=7.2cm]{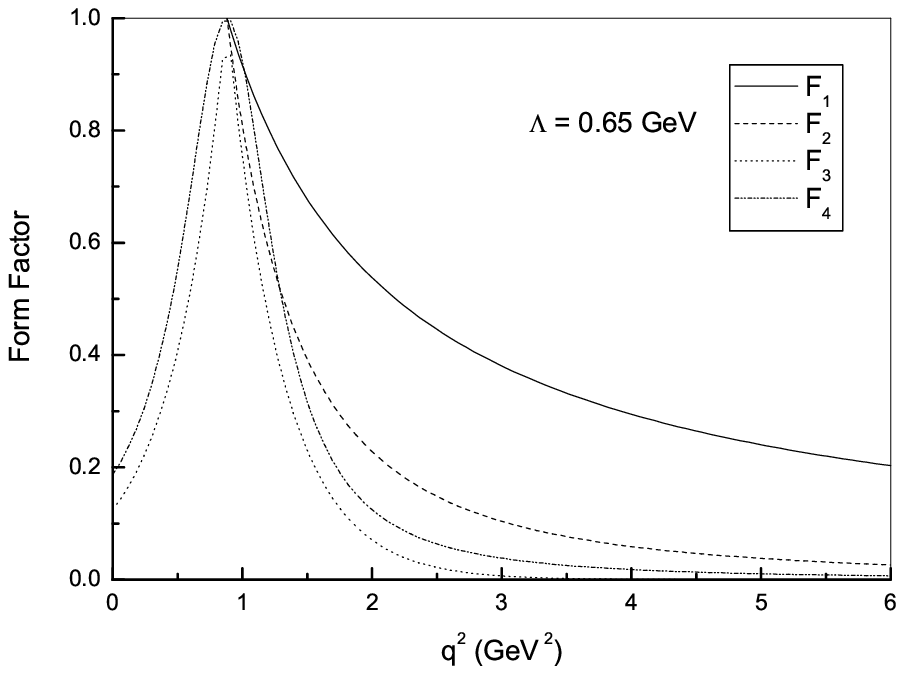}
\includegraphics[width=7.2cm]{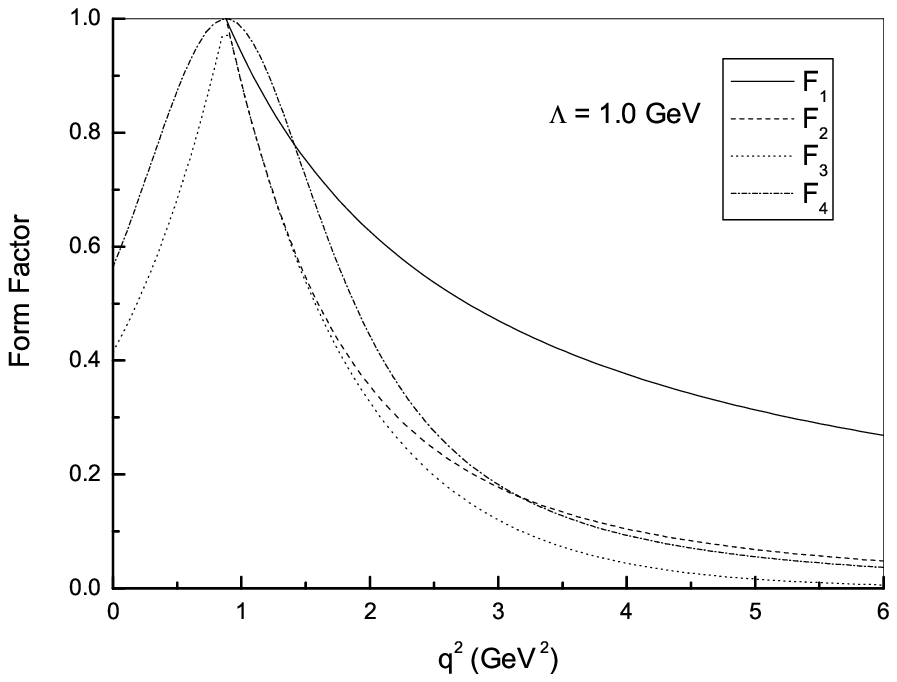}
\includegraphics[width=7.2cm]{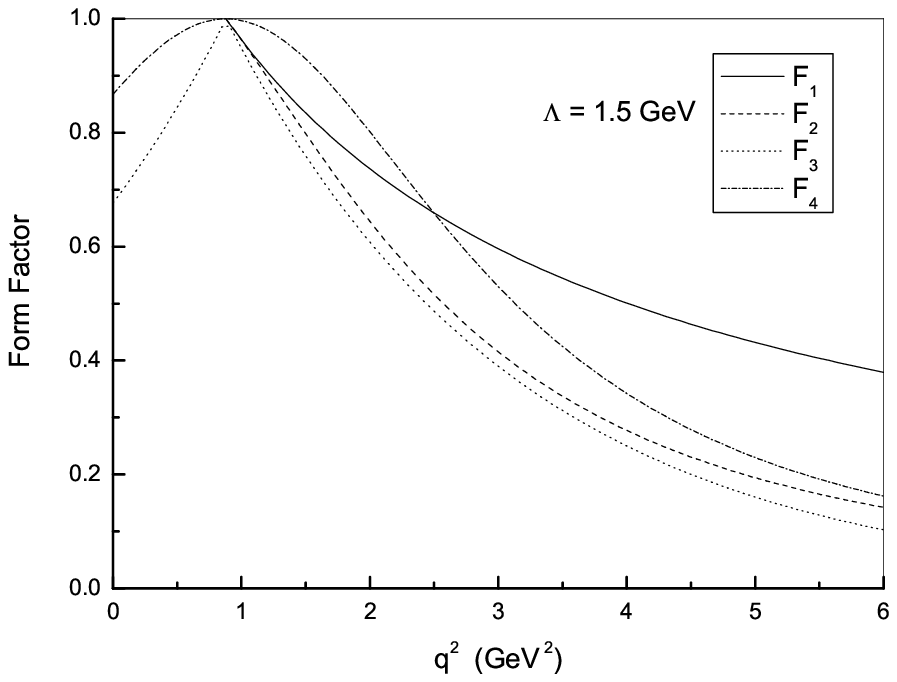}
\includegraphics[width=7.2cm]{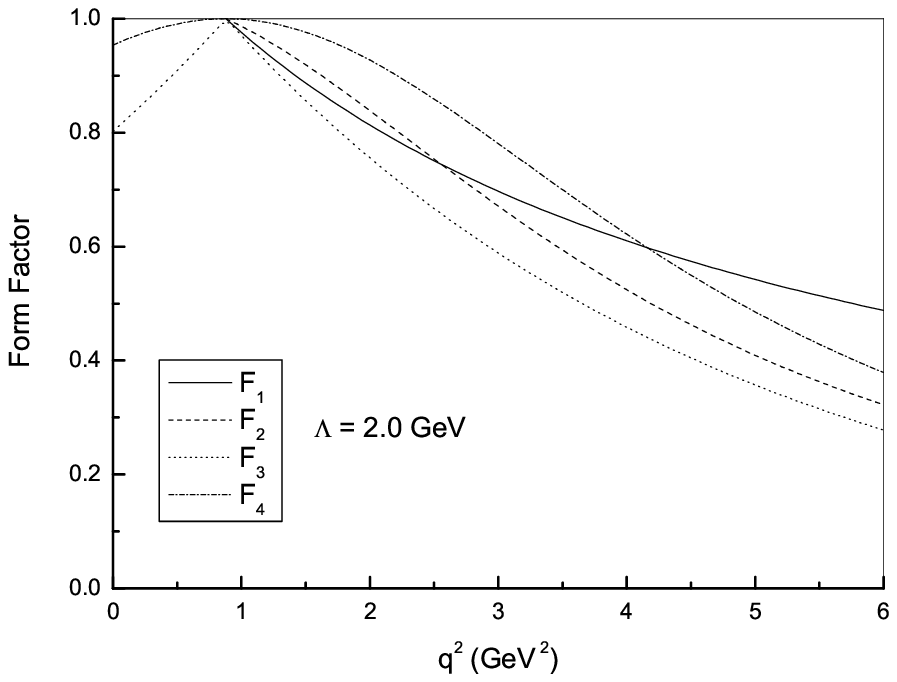}
\caption{\small {The
momentum-dependence of form factors $F_1 \sim F_4$ with the same
value of $\Lambda$.}}
\label{fig:FormFactor-Lambda}
\end{center}
\end{figure}

To give readers a comprehensive idea of various form factors, we
plot them with $\Lambda=0.65$, $1.0$, $1.5$ and $2.0 ~GeV$ in
figs.\ref{fig:FormFactor-FF} and
\ref{fig:FormFactor-Lambda},respectively.
Fig.\ref{fig:FormFactor-FF} shows that the $\Lambda$-dependence of
the form factors given above. The common feature of these form
factors is that their high momentum transfer part is even more
reduced when $\Lambda$ value becomes smaller. The
momentum-dependent behaviors of these form factors are quite
different with different $\Lambda$ values.
Fig.\ref{fig:FormFactor-Lambda} presents the momentum-dependence
of various form factors. The momentum-dependence of form factors
$F_2$, $F_3$ and $F_4$ are very sensitive to the $\Lambda$ value,
but $F_1$'s not. When the value of $\Lambda$ becomes smaller, the
difference among various form factors is more pronounced. For
instance, when $\Lambda=0.65~GeV$, with increase $q^2$, $F_3$
reduces much more than $F_1$ does. Since in the decay processes
considered in this paper, the intermediate nucleon is off-shell,
introducing an off-shell form factor would suppress  the off-shell
effect of the nucleon, and the form of the form factor and the
value of $\Lambda$ would affect the decay amplitude. Therefore,
studying the hadronic vertex form factor can provide a constraint
to the data analysis. Moreover, the results of the data fitting
can also help us to choose a proper form factor for the
$J/\psi\to N \bar{N} M$ investigation.

%%%%%%%%%%%%%%%%%%% Fig. 5 %%%%%%%%%%%%%%%%%%%
%%%%%%%%%%%%%%%%%%%%%%%%%% Fig. 6 %%%%%%%%%%%%%%%%%%

Suppose the form factors with same form are  applied on both vertices of the
considered decay diagram, respectively. After adding form factors,
Eqs. (\ref{ps-on}), (\ref{ps-off}) and (\ref{off-pv})can be
re-written as
\begin{eqnarray}
{\cal{M}}^{\prime}_{PS} &=& i g_{N\bar{N} \pi}  \bar{u}(p)
\gamma_5~ \left[~ F_M \left( \frac{\slash{\hskip -2.0mm}k
~\slash{\hskip -2.0mm} \epsilon}{2p\cdot k +k^2} F^2(q^2)
-\frac{\slash{\hskip -2.0mm} \epsilon ~\slash{\hskip
-2.0mm}k}{2p^{\prime}\cdot k +k^2} F^2(q^{\prime 2})
\right) \right.\nonumber \\
&~& \left. +\frac{F_0}{m} ~\slash{\hskip -2.0mm}k \left( \frac{p
\cdot \epsilon}{2p^{\prime}\cdot k +k^2} F^2(q^{\prime 2})
-\frac{p^{\prime} \cdot \epsilon}{2p\cdot k +k^2} F^2(q^2) \right)
\right]~v(p^{\prime}),
\end{eqnarray}
and
\begin{equation}
{\cal{M}}^{\prime}_{PV}=\frac{ig_{N\bar N \pi}}{2m} \bar{u}(p) \{
F_M [F^2(q^2)-F^2(q^{\prime 2}) ] ~\slash{\hskip -2.0mm} \epsilon
+ \frac{F_0}{m} [F^2(q^{\prime 2}) (p \cdot \epsilon) -F^2(q^2)
(p^{\prime} \cdot \epsilon) ] \} \gamma_5 v(p^{\prime})
+{\cal{M}}^{\prime}_{PS},
\end{equation}
respectively. Form factor $F(q^2)$ can be chosen to be one of the
forms in Eqs. (\ref{F1}) -- (\ref{F5}).

In order to show how sensitive the decay BR to the form of the
form factor and the value of $\Lambda$, we take $\Lambda =$ 0.65,
1.0, 1.5 and 2.0~GeV and $|F_0|/|F_M| = 0.12 \pm 0.08$, and
calculate the BR of $\Gamma(J/\psi \to p \bar{p} \pi^0)/
{\Gamma(J/\psi \to p \bar{p} )} $ with all kinds of form
factors shown above. The results are tabulated in Table
\ref{tab:pi-FF}, where the numbers in parentheses are
corresponding to the lower and  the upper limits of
$|F_0|/|F_M|$, respectively.
%
%%%%%%%%%%%%%%%%%%%%%%%%%%%% Table 1 %%%%%%%%%%%%%%%%%%%%%%
{\def\baselinestretch{1.2}
\begin{table}[htb]
\caption
{\small {The BR of $\Gamma(J/\psi \to p \bar{p} \pi^0)/
\Gamma(J/\psi \to p \bar{p} )$ (\%) with various form factors.}}
%\vspace{-0.3cm}
\label{tab:pi-FF}
\begin{center}
\begin{footnotesize}
\begin{tabular}{|c|c|c|c|c|c|}
\hline
F.F. &  $\pi$N & $\Lambda=0.65~GeV$ & $\Lambda=1.0~GeV$
& $\Lambda=1.5~GeV$ & $\Lambda=2.0~GeV$ \\
&  coupling & & & &  \\
\hline
$F_1$ &  PS &  3.95(3.73$\sim$4.18) & 6.81(6.45$\sim$7.20)
& 12.69(12.05$\sim$13.38) & 19.35(18.40$\sim$20.37)\\
\cline{2-6}
&  PV &  2.79(2.77$\sim$2.82) & 5.04(5.01$\sim$5.07)
& 9.96(9.91$\sim$9.98) & 15.89(15.83$\sim$15.91)\\
\hline
$F_2$ &  PS & 0.34(0.32$\sim$0.37) & 1.23(1.15$\sim$1.31)
& 7.21(6.82$\sim$7.64) &  19.64(18.66$\sim$20.71)\\
\cline{2-6}
& PV & 0.20(0.19$\sim$0.21) & 0.76(0.75$\sim$0.78)
&  5.07(5.02$\sim$5.11) &  15.50(15.45$\sim$15.53)\\
\hline
$F_3$ &  PS & 0.07(0.06$\sim$0.07) & 1.09(1.02$\sim$1.16)
&  5.83(5.51$\sim$6.18) &  13.29(12.61$\sim$14.02)\\
\cline{2-6}
& PV & 0.04(0.03 $\sim$ 0.04) & 0.66(0.64$\sim$0.68)
& 4.07(4.03$\sim$4.10) & 10.23(10.18$\sim$10.25)\\
\hline
$F_4$ & PS & 0.23(0.22$\sim$0.25) & 3.35(3.15$\sim$3.58)
& 15.03(14.23$\sim$15.89) & 29.70(29.68$\sim$31.26)\\
\cline{2-6}
& PV & 0.13(0.12$\sim$0.13) & 2.08(2.04$\sim$2.14)
& 10.98(10.92$\sim$11.04) & 24.30(24.23$\sim$24.31)\\
\hline
$F_5$ & PS  & 2.39(2.25$\sim$2.54) & 10.25(9.71$\sim$10.83)
&  23.91(22.75$\sim$25.16) &  34.01(32.40$\sim$35.72)\\
\cline{2-6}
& PV
& 2.33(2.16$\sim$2.81) & 9.28(9.37$\sim$9.33)
&  21.98(22.12$\sim$21.85) &  31.57(31.63$\sim$31.44)\\
\hline
\end{tabular}
\end{footnotesize}
\end{center}
\end{table}}
From Table \ref{tab:pi-FF}, one sees that no matter
which form factor is employed, the difference between
$\Gamma_{PV}(J/\psi \to p \bar{p} \pi^0)$ and
$\Gamma_{PS}(J/\psi \to p \bar{p} \pi^0)$ is generally
larger than that in the without form factor case. For instance,
when $\Lambda=1.0~GeV$,
\begin{equation}
\frac{\Gamma_{PV}(J/\psi \to p \bar{p}\pi^0)}{\Gamma_{PS}(J/\psi
\to p \bar{p} \pi^0)}  =\left\{
\begin{array}{ll}
0.74   & ~~for ~~F_1~,\\
0.618  & ~~for ~~F_2~,\\
0.606  & ~~for ~~F_3~,\\
0.621  & ~~for ~~F_4~,\\
0.905  & ~~for ~~F_5~,\\
\end{array}
\right.
%\label{eq:piBRPS1}
\end{equation}
and the corresponding ratio without form factors is
\begin{equation}
\frac{\Gamma_{PV} (J/\psi \to p \bar p \pi^0)}{\Gamma_{PS} (J/\psi
\to p \bar p \pi^0)} \simeq 0.940~. \label{eq:PV:PS-NoFF}
\end{equation}

It means that introduced form factor suppresses the contribution
at the large momentum transfer region, and consequently, enlarges
the off-shell effects in the PS-PS and PS-PV coupling cases in
different extent. For a specific form factor, when the $\Lambda$
value reduces, the curve of the form factor bends towards the
lower momentum direction. It further suppresses the contribution
at the high momentum transfer region, enlarges the off-shell
effect and reduces the nucleon-pole contribution. For instance,
with form factor $F_5$, when $\Lambda$ reduces from 2.0~GeV to 0.65~GeV,
BR $\frac{\Gamma_{PV}(J/\psi \to p \bar{p}\pi^0)}{\Gamma(J/\psi \to
p \bar{p})}$ declines from 31.57\% to 2.33\%. The $\Lambda$
dependence of the BR differs in different form factor cases. For
the same amount of $\Lambda$ value change, BR with $F_1$ decreases
about 1/5, but BR with $F_3$ drops about 1/256. The BR with a
larger $\Lambda$ value is more pronounced, and the nucleon-pole
contribution is important. On the contrary, the nucleon-pole
contribution by using a small $\Lambda$ value can be ignored.

The BR with different form factor but the same $\Lambda$ value is
quite different. For example, when $\Lambda=0.65~GeV$, the BR with
$F_3$ and $F_1$ in the PS-PV coupling case are about 0.04\% and
2.79\%, respectively. The later is about 70 times larger than the
former one. Anyway, these BRs are negligibly compared with the
data of 51\%. However, when $\Lambda$ is large, the difference
between different form factor cases becomes very small,
consequently, the contribution from the high momentum part is not
much suppressed, the resultant BRs in both PS-PS and PS-PV cases
are close to each other and are comparable with the data. For
instance, when $\Lambda=2.0~GeV$, the maximum range of BR change is
from 10.23\% to 34.01\%.

Since it is not sure which form factor is suitable for considered
decay processes, we cannot conclude whether the nucleon-pole is
dominantly responsible for the BR of $J/\psi \to N \bar{N}\pi$
decay. Now, we would show how much the nucleon-pole diagram would
contribute to the BR of $J/\psi \to N \bar{N}\pi$, when a form
factor used in a similar processes is adopted. It should be
mentioned that although all the form factors shown above are the
$NN\pi$ vertex form factor, the particle that the momentum
variable corresponds to is different case by case. In the N-N
interaction, the form factor is $\pi$-momentum dependent, and in
the $\pi$-N interaction, the pion photoproduction, and the $J/\psi
\to N \bar{N}\pi$ decay, it is intermediate-nucleon-momentum
dependent. Only in the case that the form factor depends on the
four-momenta of three interacting particles \cite{Gross93}, a
unified form factor with the same proper parameter $\Lambda$ can
possibly be applied to all mentioned processes. We summarize part
of form factors whose momentum dependence is similar to the
$J/\psi \to N \bar{N}\pi$ decay and whose $\Lambda$ value has
well been determined by the $\pi$-N scattering or the pion
photoproduction in Table \ref{tab:parameter}.
%
%%%%%%%%%%%%%%%%%%%%%% Table 2 %%%%%%%%%%%%%%%%%%%%%%%%%%%
\begin{small}
\begin{table}[htb]
\caption{\small Frequently used $\pi NN$ form factor in literatures.}
\label{tab:parameter}
\begin{center}
\begin{tabular}{clcll}
\hline \hline
$\pi N$ Coupling & ~~~Coupling Constant & ~~~F.F. & ~~~$\Lambda$(cut-off) & Ref.\\
\hline
PV & ~~~$f^2_{\pi NN}/4\pi =0.0778$ & ~~~$F_1(q^2)$ & ~~~$1350~MeV$ & \cite{Schutz94} \\
%\hline
PV & ~~~$g^2_{\pi NN}/4\pi =14.3$ & ~~~$F_4(q^2)$ & ~~~$1116.6~MeV$ & \cite{Pearce91} \\
%\hline
PV & ~~~$f^2_{\pi NN}/4\pi =0.0778$ & ~~~$F_4(q^2)$ & ~~~$1200~MeV$ & \cite{Schutz94} \\
%\hline
PS & ~~~$g^2_{\pi NN}/4\pi =14$ & ~~~$F_5(q^2)$ & ~~~$1000~MeV$ & \cite{Lee1} \\
\hline \hline
\end{tabular}
\end{center}
\end{table}
\end{small}

With these form factors, we re-calculate the BR of
$\Gamma(J/\psi \to p \bar{p} \pi^0)/ \Gamma(J/\psi \to p \bar{p}
)$. The resultant BRs with $|F_0|/|F_M|=0.12$ are:
\begin{equation}
\frac{\Gamma(J/\psi \to p \bar{p}\pi^0)}{\Gamma(J/\psi \to p
\bar{p})}  =\left\{
\begin{array}{ll}
0.0808   & ~~for ~~F_1~,\\
0.0465   & ~~for ~~F_4~,\\
0.0967   & ~~for ~~F_5~.\\
\end{array}
\right. \label{eq:pi-ppoleWithParameter}
\end{equation}
And those with $|F_0|/|F_M|=1.04$ are:
\begin{equation}
\frac{\Gamma(J/\psi \to p \bar{p}\pi^0)}{\Gamma(J/\psi \to p
\bar{p})}  =\left\{
\begin{array}{ll}
0.0735   & ~~for ~~F_1~,\\
0.0524   & ~~for ~~F_4~,\\
0.1636   & ~~for ~~F_5~.\\
\end{array}
\right. \label{eq:pi-ppoleWithParameter}
\end{equation}
Comparing with the results in Table \ref{tab:pi-FF}, one finds
that the resultant BRs do not differ as much large as those in
Table \ref{tab:pi-FF}, but there is still visible difference.

\begin{figure}[htbp]
%\vspace{0.6cm}
\begin{center}
\includegraphics[width=9.0cm]{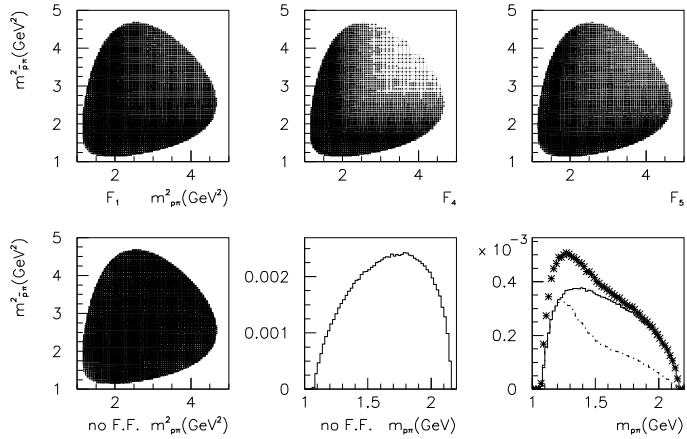}
\caption {\small {The Dalitz plot
and the invariant $p\pi^{0}$ mass distribution of the $J/\psi \to
p \bar{p} \pi^0$ decay with the form factors $F_1$, $F_4$, and
$F_5$. The solid curve, the dotted-dashed curve and the stared
curve in the invariant mass distribution figure correspond to the
form factors $F_1$, $F_4$, and $F_5$, respectively.}}
\label{fig:pi-MC-FF}
\end{center}
\end{figure}

Although the magnitude of BR is much smaller than data, it can
still be used to select a proper form factor for the $J/\psi \to N
\bar{N} \pi$ decay. To fulfil this goal, in terms of the Monte
Carlo simulation, we calculate the Dalitz plot and the invariant
$p\pi^{0}$ mass distribution of the $J/\psi \to p \bar{p} \pi^0$
decay with the form factors $F_1$, $F_4$, and $F_5$, respectively.
They are shown in Fig. \ref{fig:pi-MC-FF}.
%
%%%%%%%%%%%%%%%%%%%%%%% Fig. 7 %%%%%%%%%%%%%%%%%%%%%%%%%%
Comparing these figures with the data, one should be able to find
out a most suitable  form factor for the $J/\psi \to p \bar{p}
\pi^0$ decay.

\vspace{0.5cm}

\subsection{Nucleon-pole contributions in
$J/\psi \to p \bar{p} \eta$ and $J/\psi \to p \bar{p}
\eta^{\prime}$ decays}

%%%%%%%%%%%%%%%%%%%%%%% Fig. 8 %%%%%%%%%%%%%%%%%%%%%%%%%%
\begin{figure}[htbp]
\begin{center}
\epsfxsize=3.0in
\epsfbox{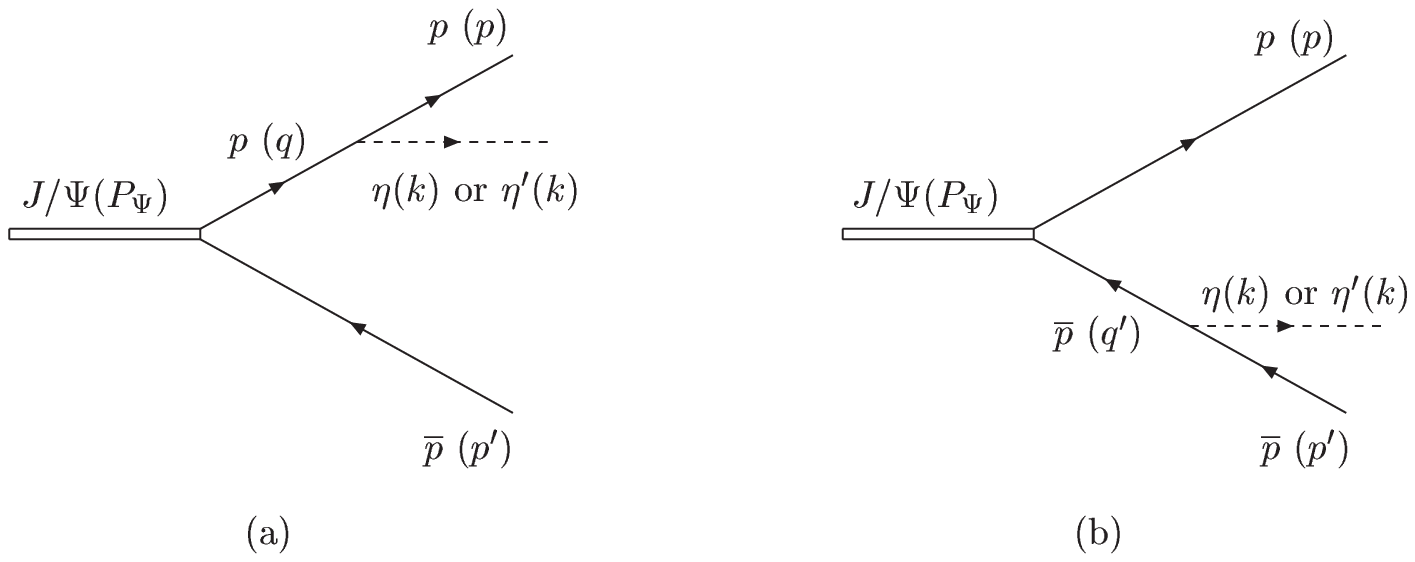}
\caption{\small
{Proton-pole diagrams for $J/\psi \to  p \bar{p} \eta$ and
$J/\psi \to  p \bar{p} \eta^{\prime}$  decays.}}
\label{eta}
\end{center}
\end{figure}

The corresponding Feynman diagrams for $J/\psi \to p \bar
p \eta$ and $p \bar p \eta^{\prime}$ decays are shown in Fig. \ref{eta}.
The same formulae for $J/\psi \to p \bar{p} \pi^0$ decay can be
applied to the $J/\psi \to p \bar{p} \eta$ and $J/\psi \to p
\bar{p} \eta^{\prime}$ decays, except replacing $g_{N \bar N \pi}$
with $g_{N \bar N \eta}$ or $g_{N \bar N \eta^{\prime}}$, because
$\pi$, $\eta$ and $\eta^{\prime}$ are all pseudoscalar mesons.
 The values of $g_{N \bar N \eta}$ and $g_{N \bar N \eta^{\prime}}$ can
be chosen according to following relations \cite{Sinha}:
\begin{equation} {(g_{\eta NN} / g_{
\pi NN})}^2 = 3.90625 \times 10^{-3} ,~~~~ {(g_{
\eta^{\prime}NN} / g_{\pi NN})}^2 = 2.5 \times 10^{-3}~.
\label{eq:etaNCoupling}
\end{equation}

And we take $|F_0|/|F_M| = 0, 0.12, 0.74$ and $1.04$ in our calculation.
The decay ratios of $\Gamma(J/\psi \to p \bar p \eta) /
\Gamma(J/\psi \to p \bar p)$ and $\Gamma(J/\psi
\to p \bar p \eta^{\prime}) / \Gamma(J/\psi \to p
\bar p)$ from the proton-pole contribution without considering
form factors are:
\begin{equation}
\frac{\Gamma_{PS}(J/\psi \to p \bar p \eta)}{\Gamma(J/\psi \to p
\bar{p})}  =\left\{
\begin{array}{ll}
5.48 \times 10^{-4}  &~ ~for ~|F_0|/|F_M|~=0~,\\
5.63 \times 10^{-4}  &~ ~for ~|F_0|/|F_M|~=0.12~,\\
6.91 \times 10^{-4}  & ~~for ~|F_0|/|F_M|~=0.74~,\\
8.19 \times 10^{-4}  & ~~for ~|F_0|/|F_M|~=1.04~,\\
\end{array}
\right.
\label{eq:etaBRPS}
\end{equation}
\begin{equation}
\frac{\Gamma_{PV}(J/\psi \to p \bar p \eta)}{\Gamma(J/\psi \to p
\bar{p})}  =\left\{
\begin{array}{ll}
5.48 \times 10^{-4}  &~ ~for ~|F_0|/|F_M|~=0~,\\
5.26 \times 10^{-4}  &~ ~for ~|F_0|/|F_M|~=0.12~,\\
4.69 \times 10^{-4}  & ~~for ~|F_0|/|F_M|~=0.74~,\\
4.12 \times 10^{-4}  & ~~for ~|F_0|/|F_M|~=1.04~,\\
\end{array}
\right.
\label{eq:etaBRPV}
\end{equation}
and
\begin{equation}
\frac{\Gamma_{PS}(J/\psi \to p \bar p \eta')}{\Gamma(J/\psi \to p
\bar{p})}  =\left\{
\begin{array}{ll}
1.93 \times 10^{-5}  &~ ~for ~|F_0|/|F_M|~=0~,\\
1.99 \times 10^{-5}  &~ ~for ~|F_0|/|F_M|~=0.12~,\\
2.45 \times 10^{-5}  & ~~for ~|F_0|/|F_M|~=0.74~,\\
2.91 \times 10^{-5}  & ~~for ~|F_0|/|F_M|~=1.04~,\\
\end{array}
\right.
\label{eq:etaprimeBRPS}
\end{equation}
\begin{equation}
\frac{\Gamma_{PV}(J/\psi \to p \bar p \eta')}{\Gamma(J/\psi \to p
\bar{p})}  =\left\{
\begin{array}{ll}
1.93 \times 10^{-5}  &~ ~for ~|F_0|/|F_M|~=0~,\\
1.87 \times 10^{-5}  &~ ~for ~|F_0|/|F_M|~=0.12~,\\
1.70 \times 10^{-5}  & ~~for ~|F_0|/|F_M|~=0.74~,\\
1.54 \times 10^{-5}  & ~~for ~|F_0|/|F_M|~=1.04~.\\
\end{array}
\right.
\label{eq:etaprimeBRPV}
\end{equation}

Again, the difference of BRs between PS-PS and PS-PV couplings
descents when the ratio of $|F_0|/|F_M|$ declines. Comparing with
the empirical data of $\Gamma(J/\psi \to p \bar p \eta) /
\Gamma(J/\psi \to p \bar p) = 0.98 \pm 0.09$ and
$\Gamma(J/\psi \to p \bar p \eta') /
\Gamma(J/\psi \to p \bar p) = 0.42 \pm 0.19$ \cite{PDG2000}, one
finds that the calculated BRs are all smaller than 0.1\% of the
data. This is because that in these two decays, the intermediate
nucleon is largely off-shell. We also tabulate the BRs of
$\frac{\Gamma(J/\psi \to p \bar p \eta)} {\Gamma(J/\psi
\to p \bar p)}$ and $\frac{\Gamma(J/\psi \to p
\bar p \eta^{\prime})} {\Gamma(J/\psi \to p \bar p)}$ with
various $\pi NN$ form factors in
Tables \ref{tab:eta-FF} and \ref{tab:etaprime-FF}, respectively.
Because the form factor further reduces the proton-pole
contribution at the high momentum region, the resultant BRs are
very small. Therefore, in analyzing the $J/\psi \to p
\bar{p} \eta$ and $J/\psi \to p \bar{p} \eta^{\prime}$
data, the proton-pole contribution can safely be  ignored. The
main contributor for such decays must be some other diagrams, for
instance, the $N^*$-pole diagram.

%%%%%%%%%%%%%%%%%%%%%%% Table 3 %%%%%%%%%%%%%%%%%%%%%%%%%%
\begin{table}[htb]
\caption{\small {Branching ratio $\Gamma(J/\psi \to p
\bar{p} \eta)/ {\Gamma(J/\psi \to p \bar{p} )} $ with form
factors, with $|F_0|/|F_M|=0.12.$}}
%\vspace{-0.3cm}
\label{tab:eta-FF}
\begin{center}
\begin{small}
\begin{tabular}{|c|c|c|c|c|c|}
\hline
F.F. & $\pi \eta$ &$\Lambda=0.65GeV$ & $\Lambda=1.0GeV$
& $\Lambda=1.5GeV$ & $\Lambda=2.0GeV$ \\
& coupling & & & &  \\
\hline
$F_1$& PS & $9.61 \times 10^{-6}$ & $2.24 \times 10^{-5}$
& $5.92 \times 10^{-5}$ & $1.14 \times 10^{-4}$\\
\cline{2-6}
& PV & $7.26 \times 10^{-6}$ & $1.74 \times 10^{-5}$
& $4.77 \times 10^{-5}$ & $9.49 \times 10^{-5}$\\
\hline
$F_2$ & PS & $6.34 \times 10^{-8}$ &$5.04 \times 10^{-7}$
& $1.31 \times 10^{-5}$ & $8.90 \times 10^{-5}$\\
\cline{2-6}
& PV &$4.15 \times 10^{-8}$ &$3.35 \times 10^{-7}$
& $9.26 \times 10^{-6}$ & $6.91 \times 10^{-5}$\\
\hline
$F_3$ & PS & $2.63 \times 10^{-11}$ &$1.54 \times 10^{-7}$
& $1.00 \times 10^{-5}$ & $5.31 \times 10^{-5}$\\
\cline{2-6}
& PV &$1.63 \times 10^{-11}$ &$1.01 \times 10^{-7}$
& $7.02 \times 10^{-6}$ & $4.09 \times 10^{-5}$\\
\hline
$F_4$ & PS & $1.77 \times 10^{-9}$ &$6.77 \times 10^{-7}$
& $3.40 \times 10^{-5}$ & $1.63 \times 10^{-4}$\\
\cline{2-6}
& PV &$1.11 \times 10^{-9}$ &$4.37 \times 10^{-7}$
& $2.39 \times 10^{-5}$ & $1.29 \times 10^{-4}$\\
\hline
$F_5$& PS & $7.06 \times 10^{-7}$ &$2.64 \times 10^{-5}$
& $1.38 \times 10^{-4}$ & $2.52 \times 10^{-4}$\\
\cline{2-6}
& PV &$3.66 \times 10^{-6}$ &$2.19 \times 10^{-5}$
& $1.22 \times 10^{-4}$ & $2.29 \times 10^{-4}$\\
\hline
\end{tabular}
\end{small}
\end{center}
\end{table}
%
%%%%%%%%%%%%%%%%%%%%%%% Table 4 %%%%%%%%%%%%%%%%%%%%%%%%%%
\begin{table}[htb]
\caption{\small {Branching ratio $\Gamma(J/\psi \to
p \bar{p} \eta^{\prime})/ {\Gamma(J/\psi \to p \bar{p} )}
$ with form factors, with $|F_0|/|F_M|=0.12.$}}
%\vspace{-0.3cm}
\label{tab:etaprime-FF}
\begin{center}
\begin{small}
\begin{tabular}{|c|c|c|c|c|c|}
\hline F.F. & $\pi \eta^{\prime}$ &$\Lambda=0.65GeV$
& $\Lambda=1.0GeV$ & $\Lambda=1.5GeV$ & $\Lambda=2.0GeV$ \\
& coupling & & & &  \\
\hline
$F_1$& PS & $1.35 \times 10^{-7}$ & $3.61 \times 10^{-7}$
& $1.16 \times 10^{-6}$ & $2.59 \times 10^{-6}$\\
\cline{2-6}
& PV & $1.14 \times 10^{-7}$ & $3.09 \times 10^{-7}$
& $1.01 \times 10^{-6}$ & $2.28 \times 10^{-6}$\\
\hline
$F_2$ & PS & $1.99 \times 10^{-10}$ &$1.96 \times 10^{-9}$
& $9.94 \times 10^{-8}$ & $1.33 \times 10^{-6}$\\
\cline{2-6}
& PV &$1.53 \times 10^{-10}$ &$1.52 \times 10^{-9}$
& $7.93 \times 10^{-8}$ & $1.11 \times 10^{-6}$\\
\hline
$F_3$ & PS & $4.78 \times 10^{-18}$ &$5.99 \times 10^{-11}$
& $6.40 \times 10^{-8}$ & $7.76 \times 10^{-7}$\\
\cline{2-6}
& PV &$3.62 \times 10^{-18}$ &$4.16 \times 10^{-11}$
& $4.99 \times 10^{-8}$ & $6.50 \times 10^{-7}$\\
\hline
$F_4$ & PS & $1.70 \times 10^{-12}$ &$1.22 \times 10^{-9}$
& $2.28 \times 10^{-7}$ & $2.63 \times 10^{-6}$\\
\cline{2-6}
& PV &$1.26 \times 10^{-12}$ &$9.15 \times 10^{-10}$
& $1.79 \times 10^{-7}$ & $2.21 \times 10^{-6}$\\
\hline
$F_5$& PS & $8.13 \times 10^{-10}$ &$2.67 \times 10^{-7}$
& $2.91 \times 10^{-6}$ & $6.75 \times 10^{-6}$\\
\cline{2-6}
& PV &$1.23 \times 10^{-7}$ &$2.32 \times 10^{-7}$
& $2.50 \times 10^{-6}$ & $6.05 \times 10^{-6}$\\
\hline
\end{tabular}
\end{small}
\end{center}
\end{table}

\vspace{0.5cm}

\section{Nucleon-pole contribution in the $J/\psi \to p
\bar{p} \omega $ decay}

\begin{figure}[htbp]
\vspace{0.6cm}
\begin{center}
\includegraphics[width=8.5cm]{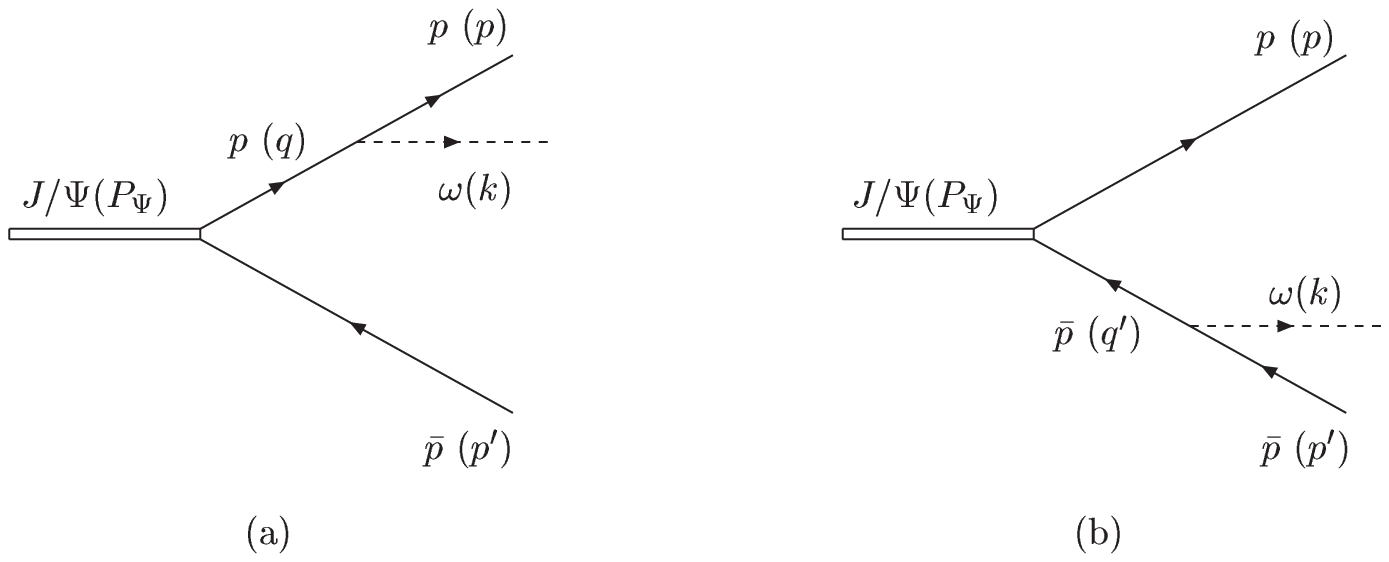}
\caption {\small {Proton-pole diagram for $J/\psi \to p \bar{p}
\omega$ decay.}}
\label{fig:omega-ppole}
\end{center}
\end{figure}

The nucleon-pole diagram in the $J/\psi \to p \bar{p}
\omega $ decay are shown in Fig.\ref{fig:omega-ppole},
%
%%%%%%%%%%%%%%%%%%%%%%% Fig. 9 %%%%%%%%%%%%%%%%%%%%%%%%%%
where the variables in brackets are four-momenta of corresponding
particles. The mass of $\omega$ meson is 781.94MeV. Due to heavy
mass of $\omega$, the intermediate nucleon in
Fig.\ref{fig:omega-ppole} must be far from the mass shell.

The $NN\omega$ interaction can be written as \cite{Sinha}
\begin{equation}
H_{\omega NN}=g_{\omega NN}\bar N (x)\gamma^{\alpha}N(x)\omega_{\alpha}(x)
+i \frac{1}{4m}f_{\omega NN}\bar{N}(x)[\gamma^{\mu},\gamma^{\nu}]
N(x)\partial_{\mu}\omega_{\nu}(x)~.
\label{eq:H-WNN}
\end{equation}
The vector coupling constant $g_{\omega NN}$ and tensor coupling
constant $f_{\omega NN}$ are:
\begin{equation}
g_{\omega pp}^{2}/4 \pi\simeq6.3~,
\end{equation}
\begin{equation}
f_{\omega pp}=(\mu_{p} + \mu_{n})g_{\omega pp}~,
\end{equation}
respectively, with the anomalous magnetic moments of proton and
neutron
\begin{equation}
\mu_{p}=1.7928 \mu_{N}~,~~~~~~~\mu_{n}=-1.9131 \mu_{N}~,
\end{equation}
respectively. A simple manipulation gives
\begin{equation}
f_{\omega pp}\simeq -0.12 g_{\omega pp}~.
\end{equation}
Therefore, the $\omega NN$ interaction is mainly vector coupling.

Similar to what Y. Oh and T-S.H. Lee did \cite{Sato96, Y.Oh01} in
the vector meson photoproduction study, we only take the vector
coupling in the $J/\psi \to p \bar{p} \omega $ calculation
\begin{equation}
H_{\omega NN}^{\prime}=g_{\omega NN}\bar
N (x)\gamma^{\alpha}N(x)\omega_{\alpha}(x)~.
\label{eq:Hprime-WNN}
\end{equation}
Performing similar derivation in the $J/\psi \to p \bar{p}
\pi^0 $ case and using the properties
\begin{equation}
P_{\psi}^{\beta}\epsilon_{\beta}(P_{\psi},\lambda)=0~,
  ~~~~k^{\alpha}e_{\alpha}(k,\lambda^{\prime})=0~,
\end{equation}
where $\epsilon_{\beta}(P_{\psi}, \lambda)$ and
$e_{\alpha}(k,\lambda^{\prime})$ are polarization vectors of
$J/\psi$ and $\omega$, respectively, we get
 the total decay amplitude for Fig. \ref{fig:omega-ppole}
\begin{eqnarray}
{\cal M}
 &=&g_{\omega pp}~\bar{u}(p,s) \left\{ F_{M} \left[
~\frac{2p\cdot e + \slash{\hskip -2.0mm}e \slash{\hskip
-2.0mm}k}{2p\cdot k+k^{2}}~\slash{\hskip -2.0mm}\epsilon
-\slash{\hskip -2.0mm}\epsilon ~\frac{2p^{\prime}\cdot e +
\slash{\hskip -2.0mm}k \slash{\hskip -2.0mm}e}{2p^{\prime}\cdot
k+k^{2}}
~\right]\right.\nonumber\\
& &-\frac{F_{0}}{m}\left.\left[~ (p^{\prime} \cdot \epsilon)~
\frac{2p\cdot e+\slash{\hskip -2.0mm}e \slash{\hskip -2.0mm}k}
{2p\cdot k+k^{2}}
+(p \cdot \epsilon)~\frac{2p^{\prime}\cdot e+
\slash{\hskip -2.0mm}k \slash{\hskip -2.0mm}e}{2p^{\prime}\cdot k+k^{2}}
~ \right]\right\} v(p^{\prime},s^{\prime})~,
\label{eq:omega-M}
\end{eqnarray}
and the differential decay width by summing over possible spin states of
the initiate and final particles
\iffalse
\begin{eqnarray}
d\Gamma(J/\psi\to p\bar{p}\omega)
&=& \frac{2\pi^{4}}{2M_{\psi}}~
\overline{\left|{\cal M}\right|^{2}}~d\Phi_{3}(P_{\psi};p,p^{\prime},k)\nonumber\\
&=&(2\pi)^{4} \frac{2g^{2}_{\omega pp}}{M_{\psi}}\left[~\left|F_{M}\right|^{2}A_{1}
+ \left|F_{0}\right|^{2}A_{2} \right. \nonumber \\
&&~~~~~~~~~~~~~~~~~~~~ \left. +Re(F_{0}^{*}F_{M})A_{3}~\right]
d\Phi_{3}(P_{\psi};p, p^{\prime}, k)~,
\label{eq:omega-DecayRate}
\end{eqnarray}
\fi
\begin{equation}
d\Gamma(J/\psi\to p\bar{p}\omega) = \frac{2\pi^{4}}{2M_{\psi}}~
\overline{\left|{\cal M}\right|^{2}}~d\Phi_{3}(P_{\psi};p,p^{\prime},k)~.
\label{eq:omega-DecayRate}
\end{equation}

Taking $|F_{0}|/|F_{M}|=0$, $0.12$, $0.74$ and $1.04$, we obtain the
branching ratio
\begin{equation}
\frac{\Gamma(J/\psi \to p \bar p \omega)}{\Gamma(J/\psi \to p
\bar{p})}  =\left\{
\begin{array}{ll}
0.169  &~ ~for ~|F_0|/|F_M|~=0~,\\
0.168  &~ ~for ~|F_0|/|F_M|~=0.12~,\\
0.171  & ~~for ~|F_0|/|F_M|~=0.74~,\\
0.175  & ~~for ~|F_0|/|F_M|~=1.04~.\\
\end{array}
\right.
\label{eq:omega-Ratio}
\end{equation}
In comparison with the data of $0.61\pm 0.12$ \cite{PDG2000}
%\ref{eq:Expforomega},
one sees that without considering form
factor, the proton-pole diagram provides rather important
contribution to the width of the $J/\psi \to p \bar p \omega$
decay. As mentioned above, because $\omega$ is relative heavy, the
intermediate proton should be far from mass shell, the terms with high
power of momentum in the amplitude make the amplitude vs momentum
curve bent upward and gone apart from the normal Breit-Wigner
form. This non-physical feature of the amplitude at the high
momentum region should be suppressed by adding off-shell form
factors.

After including the form factor, the decay amplitude becomes
\begin{eqnarray}
{\cal M}^{\prime}
&=&g_{\omega pp}~\bar{u}(p,s)
\left\{ F_{M} \left[~ \frac{2p\cdot e+\slash{\hskip -2.0mm}e
\slash{\hskip -2.0mm}k}{2p\cdot k+k^{2}}~ F^{2}(q^{2})~
\slash{\hskip -2.0mm}\epsilon
-\slash{\hskip -2.0mm}\epsilon ~\frac{2p^{\prime}\cdot e +
\slash{\hskip -2.0mm}k \slash{\hskip -2.0mm}e}{2p^{\prime}\cdot k+k^{2}}
~F^{2}(q^{\prime 2})~\right]\right.\nonumber\\
& &~~~~~~~~~~~~~~~~~~~~- \frac{F_{0}}{m}\left[ ~(p^{\prime} \cdot \epsilon)
~\frac{2p\cdot e+\slash{\hskip -2.0mm}e \slash{\hskip -2.0mm}k}
{2p\cdot k+k^{2}}~ F^{2}(q^{2}) \right.
\nonumber\\
& & \left. \left. ~~~~~~~~~~~~~~~~~~~~~~~~~~~~~~~
 + (p \cdot \epsilon)~\frac{2p^{\prime}\cdot e +
 \slash{\hskip -2.0mm}k \slash{\hskip -2.0mm}e}{2p^{\prime}\cdot k+k^{2}}
 ~F^{2}(q^{\prime 2}) ~\right]\right\} v(p^{\prime},s^{\prime})~,
\label{eq:omega-M-FF}
\end{eqnarray}
where the form factor $F(q^{2})$ can be any one from
Eqs. (\ref{F1})-(\ref{F5}). Taking $|F_{0}|/|F_{M}|=0.12$
again, we obtain the BRs of $\Gamma(J/\psi\to
p\bar{p}\omega)/\Gamma(J/\psi\to p\bar{p})$. They are
tabulated in Table \ref{tab:omega-FF}.
%
%%%%%%%%%%%%%%%%%%%%%%% Table 5 %%%%%%%%%%%%%%%%%%%%%%%%%%
{\def\baselinestretch{1.4}
\begin{table}[htb]
\caption {\small {The branching ratio of $\Gamma(J/\psi
\to p \bar{p} \omega)/ {\Gamma(J/\psi \to p
\bar{p} )}$ decay with form factor.}}
\vspace{-0.2cm}
\label{tab:omega-FF}
\begin{center}
\begin{small}
\begin{tabular}{|c|c|c|c|c|}
\hline
F.F.
&  $\Lambda=0.65~GeV$
& $\Lambda=1.0~GeV$
&  $\Lambda=1.5~GeV$
& $\Lambda=2.0~GeV$ \\
\hline
$F_1$
&  1.48$\times 10^{-3}$
& 3.83$\times 10^{-3}$
& 1.16$\times 10^{-2}$
&  2.49$\times 10^{-2}$\\
\hline
$F_2$
&  3.34$\times 10^{-6}$
& 3.15$\times 10^{-5}$
& 1.35$\times 10^{-3}$
&  1.48$\times 10^{-2}$\\
\hline
$F_3$
&  5.03$\times 10^{-12}$
& 2.45$\times 10^{-6}$
& 9.40$\times 10^{-4}$
& 8.66$\times 10^{-3}$\\
\hline
$F_4$
&  3.83$\times 10^{-8}$
& 2.43$\times 10^{-5}$
&  3.27$\times 10^{-3}$
&  2.88$\times 10^{-2}$\\
\hline
$F_5$
&  2.02$\times 10^{-5}$
& 3.28$\times 10^{-3}$
& 2.86$\times 10^{-2}$
&  6.18$\times 10^{-2}$\\
\hline
\end{tabular}
\end{small}
\end{center}
\end{table}}
From this table, one finds that the proton-pole contribution is
sensitive to the form of the form factor and the value of
$\Lambda$. The most $\Lambda$-sensitive form factor is $F_3$, with
which BR changes to almost $10^{8}$ times as much
when $\Lambda$ increases from 0.65~GeV to
2.0~GeV. The most $\Lambda$-insensitive one is $F_1$, with which BR
only increases to 17 times as much. Moreover,
when $\Lambda$ is small, the BR is more sensitive to the form of
the form factor. For instance, with $\Lambda=0.65~GeV$, the
resultant BRs from various form factors have almost $10^9$ times
difference. But with  $\Lambda=2.0~GeV$, the difference is just
about 3 times. The reason is the same as that in the $J/\psi
\to p \bar{p}\pi^0 $ case. We also provide the relevant
Dalitz plot and the invariant mass distribution of $p\omega$.
The Dalitz plot in these figures shows that the contributions of
the proton-pole diagram at the high momentum region is evidently
suppressed. This agrees with our conjecture mentioned at the
beginning of this section.
%
%%%%%%%%%%%%%%%%%%%%%%% Fig. 10 %%%%%%%%%%%%%%%%%%%%%%%%%%
\begin{figure}[htbp]
\vspace{0.6cm}
\begin{center}
\includegraphics[width=10cm]{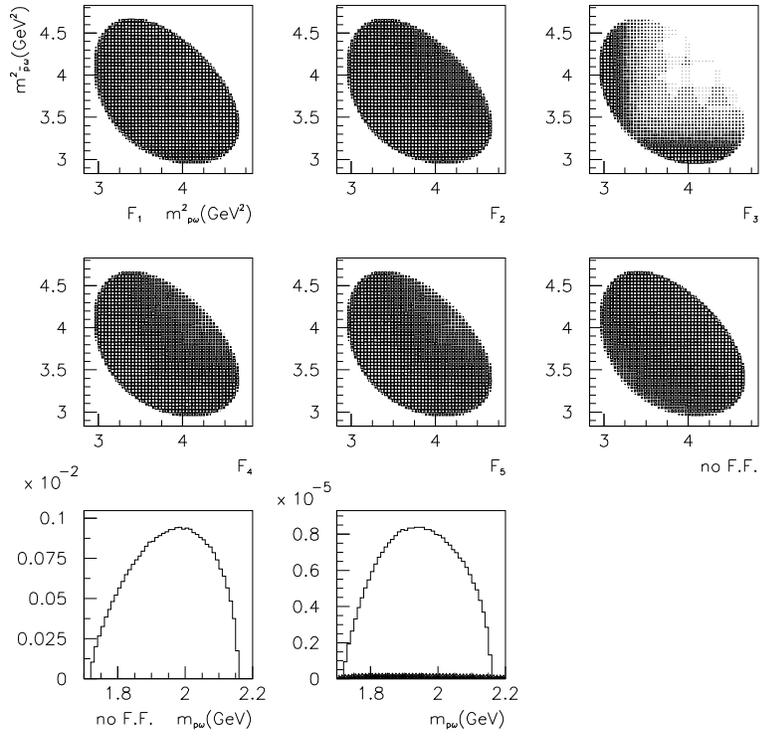}
\caption {\small{The Dalitz plot and the $p\omega$
invariant mass distribution in $J/\psi \to p \bar p \omega$ decay
(~$\Lambda =0.65 ~GeV$~). The solid, dashed, dotted, dotted-dashed
and stared curves denote the cases with $F_1$, $F_2$, $F_3$, $F_4$
and $F_5$, respectively}}
\label{fig:omega-MC-lam065}
\end{center}
\end{figure}
%
%%%%%%%%%%%%%%%%%%%%%%%% Fig. 11 %%%%%%%%%%%%%%%%%%%%%%%%%%
\begin{figure}[htbp]
\vspace{0.6cm}
\begin{center}
\includegraphics[width=10cm]{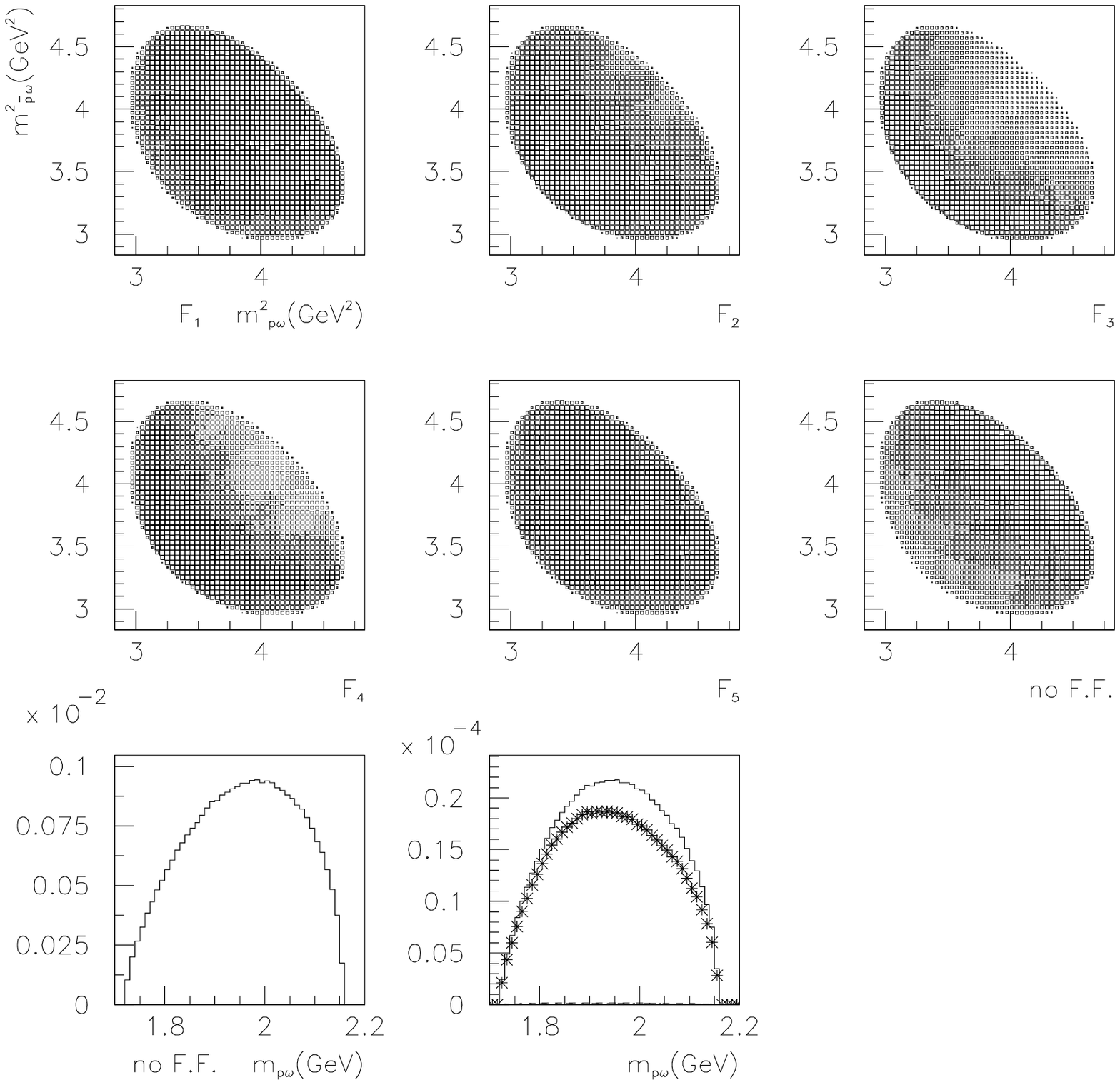}
\caption  {\small {The Dalitz plot and the $p\omega$ invariant mass
distribution in $J/\psi \to p \bar p \omega$ decay (~$\Lambda =1.0
~GeV$~). The solid, dashed, dotted, dotted-dashed and stared
curves denote the cases with $F_1$, $F_2$, $F_3$, $F_4$ and
$F_5$, respectively}}
\label{fig:omega-MC-lam100}
\end{center}
\end{figure}
%
%%%%%%%%%%%%%%%%%%%%%%%% Fig. 12 %%%%%%%%%%%%%%%%%%%%%%%%%%
\begin{figure}[htbp]
\vspace{0.6cm}
\begin{center}
\includegraphics[width=10cm]{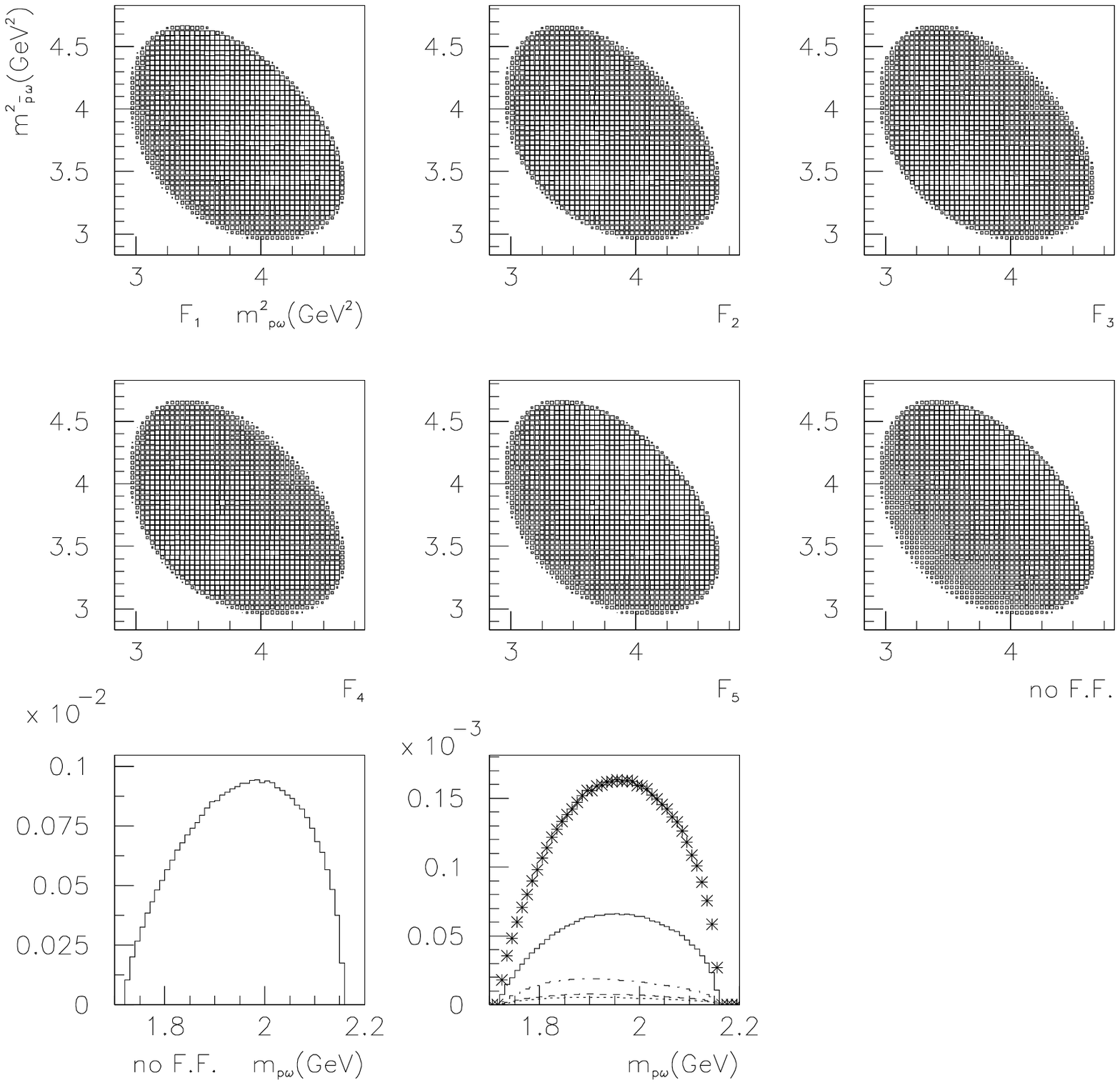}
\caption {\small {The Dalitz plot and the $p\omega$ invariant mass
distribution in $J/\psi \to p \bar p \omega$ decay (~$\Lambda =1.5
~GeV$~). The solid, dashed, dotted, dotted-dashed and stared
curves denote the cases with $F_1$, $F_2$, $F_3$, $F_4$ and
$F_5$, respectively}}
\label{fig:omega-MC-lam150}
\end{center}
\end{figure}
%
%%%%%%%%%%%%%%%%%%%%%%%% Fig. 13 %%%%%%%%%%%%%%%%%%%%%%%%%%
\begin{figure}[htbp]
\vspace{0.6cm}
\begin{center}
\includegraphics[width=10cm]{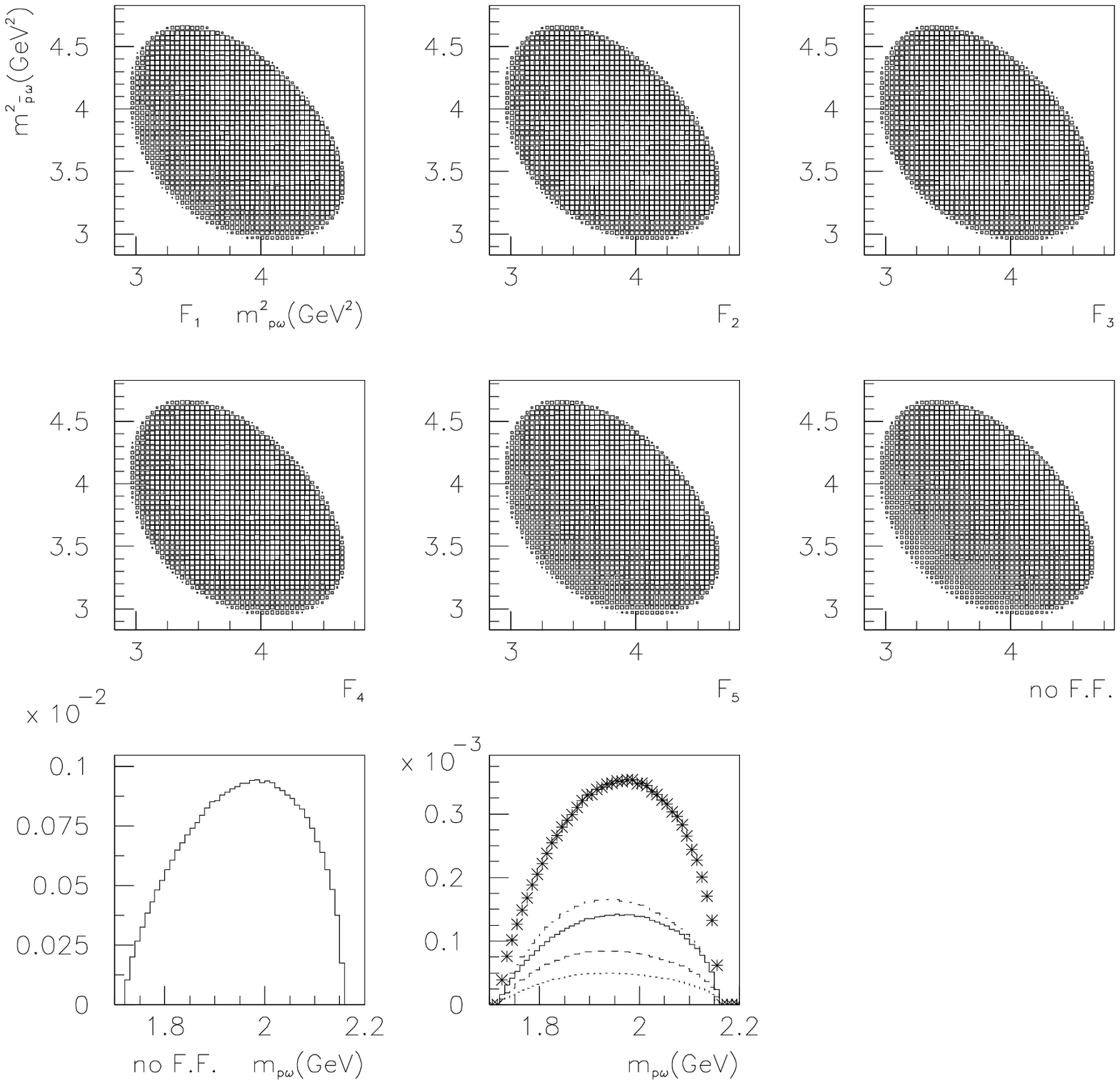}
\caption {\small {The Dalitz plot and the $p\omega$ invariant mass
distribution in $J/\psi \to p \bar p \omega$ decay (~$\Lambda =2.0
~GeV$~). The solid, dashed, dotted, dotted-dashed and stared
curves denote the cases with $F_1$, $F_2$, $F_3$, $F_4$ and
$F_5$, respectively}}
\label{fig:omega-MC-lam200}
\end{center}
\end{figure}

Furthermore, comparing with the data of
$0.61\pm 0.12$

in PDG \cite{PDG2000}, we find that the resultant BRs are
generally less than 10\% of the data. This indicates that in the
$J/\psi\to p\bar{p}\omega$ decay, the proton-pole
contribution is not so important. To explain the empirical data,
there must be certain contributions from other diagrams such as
the $N^*$-pole diagram.

\section{conclusion}

$J/\psi\to N\bar{N}M$ decay is an ideal process to study
$N^*$ spectrum. As intermediate states, nucleon and $N^*$ can all
contribute to the decay BR. In the $J/\psi\to N\bar{N}M$
decay data analysis, nucleon-pole contribution would play an
important role of background. Understanding this contribution
would enable us to get a more accurate and more reliable
information of $N^*$. In this paper, we study the nucleon-pole
contribution by employing PS-PS and PS-PV $\pi N\bar N$ vertex couplings and
various vertex form factors.

According to the equivalent theorem \cite{Chew54}, the PS-PS and
PS-PV couplings of $\pi$-N interaction are equivalent, when the
intermediate nucleon is on-shell. Namely, the decay amplitudes
with the PS-PS and PS-PV coupling vertices are exactly the same.
But, when the intermediate nucleon is off-shell, their decay
amplitudes are different. The amplitude with the PS-PS coupling
vertex keeps the same form as in the on-shell case, and the
amplitude with the PS-PV coupling vertex has an additional term
which describes a four particle contact interaction. It seems that
the PS-PV coupling contains PS-PS coupling. In fact, many authors
claimed that using PS-PV coupling only is good enough in
describing the $\pi$-N interaction and the meson photoproduction
\cite{Pearce91, Schutz94, Sato96, Peccei68, C.Lee91}. But some
authors believed that a mixed coupling
\begin{equation}
g_{\pi NN} \tau_i \left[ \lambda \gamma_5 -
(1-\lambda)\frac{\slash{\hskip -2.0mm}p- \slash{\hskip
-2.0mm}p^{\prime}}{2m} \gamma_5 \right] ~,
\end{equation}
where $\lambda$ is a mixing parameter, is more appropriate
\cite{Gross92, Gross93, Goudsmit93}. The
value of $\lambda$ can be extracted by data fitting. For an
example, Gross, Orden and Holinde \cite{Gross92} obtained
$\lambda\cong 0.22$ by fitting the N-N data in a one-boson
exchange (OBE) model, Goudsmit£¬Leisi and Mastinos found
$\lambda\cong 0.24$ by analyzing the $\pi$-N scattering data at
the tree diagram level \cite{Goudsmit93}, Gross and
Surya got $\lambda\cong 0.25$ by fitting the $\pi$-N scattering
data \cite{Gross93}. Anyway, the obtained $\lambda$ value shows
that in the mixed $NN\pi$ vetex, PS-PS coupling only occupies a
samll portion.

Because PS-PS and PS-PV couplings are not equivalent when the
intermediate nucleon is off-shell, the resultant nucleon-pole
contribution to $\Gamma(J/\psi\to p\bar{p}\pi^0)/\Gamma(J/\psi\to
p\bar{p})$ in these cases are different. The smaller the
$|F_0|/|F_M|$ ratio in the $J/\psi\to p\bar{p}$ decay is, the
closer the BRs in the PS-PS and PS-PV cases are. For instance,
when $|F_0|/|F_M|=0.12$, the mentioned difference is quite small,
and the ratio $\Gamma_{PV}(J/\psi\to
p\bar{p}\pi^0)/\Gamma_{PS}(J/\psi\to p\bar{p}\pi^0)$ is about
0.94. The resultant BR is about 0.53. In comparison with the data
of 0.51, one can claim that the proton-pole diagram  is the main
contributor to be responsible for the BR of the $J/\psi\to
p\bar{p}\pi^0$ decay.

On the other hand, hadron has its own inner structure. To be
realistic, one has to introduce  vertex form factors. After
considering the form factor, the mentioned BR difference is
enlarged. The size of the change depends on the form of the form
factor and its $\Lambda$ value. In general, the smaller the
$\Lambda$ is, the large the difference would be. When $\Lambda$ is
small, say $\Lambda=0.65~GeV$, the resultant BR of
$J/\psi\to p\bar{p}\pi^0$ depends strongly on the form of
the form factor. When the form factor takes an exponential form or
a dipole form, it highly suppresses the contribution at the high
momentum part, and consequently, the resultant BR reduces to
0.08\% or 0.4\% of the value obtained without form factor. When
$\Lambda$ is large, say $\Lambda=2.0~GeV$, all the form factors
become very similar to each other, the form factor curves do not
declined too much, and then the resultant BRs are also very
similar and are not small. They are about 20\% to 30\% of the
value without form factor.

The $\Lambda$-sensitivity of various form factors are also
different. The exponential form factor is the most sensitive one.
When $\Lambda$ value changes from 0.65~GeV to 2.0~GeV, the BR
changes about 256 times as much. But for a most-insensitive one
(monopole form factor), the change is only about 6 times as much.

Moreover, if one adopts a form factor that are frequently used in
explaining the $\pi$-N scattering and the pion photoproduction
data, the proton-pole contribution is about $10\sim 20$\% of the
$J/\psi\to p\bar{p}\pi^0$ data. Thus, the proton-pole
diagram must be accounted.

The similar results for the $J/\psi\to p\bar{p}\eta$ and
$p\bar{p}\eta^{\prime}$ are studied in the same manner.
Taking $|F_0|/|F_M|=0.12$ and
without
considering the form factor, the resultant BRs from the
proton-pole diagram are $5\times10^{-4}$ and $2\times10^{-4}$, for
the $J/\psi\to p\bar{p}\eta$ and $p\bar{p}\eta^{\prime}$
decays, respectively. In comparison with the data of $0.98$ and
$0.42$, they are all less than $0.1$\% of the data. Taking the
form factor into account, the resultant BRs are further reduced.
Therefore, in analyzing the $J/\psi\to p\bar{p}\eta$
and $p\bar{p}\eta^{\prime}$ decay data, the contribution from the
proton-pole diagram can safely be ignored.

The proton-pole diagram contribution to the $J/\psi\to
p\bar{p}\omega$ decay is analyzed too. The difference between
resultant BRs by using vector coupling and mixed coupling is only
about 3\%. Comparing with the data of 0.61, without considering
the form factor and with $|F_0|/|F_M|=0.12$, the BR obtained from the proton-pole diagram is
about 0.168, which is about 28\% of the data. When the form factor
is considered, the obtained largest BR is less than 10\% of the
data. This indicates that other diagram such as $N^*$-pole diagram may be
mainly responsible for the $J/\psi\to p\bar{p}\omega$
decay.

Finally, it is worthy to know that through $J/\psi$ decay data
fitting, it is possible to select an appropriate form factor for
the $J/\psi\to N \bar{N} M$ decay.

\vspace{1cm}

\centerline{\Large \bf ACKNOWLEDGMENTS}
\vspace{0.5cm}

The authors would like to thank Professor Rahul Sinha for his
fruitful discussion. We also thank Wei-Xing Ma for useful
discussions. This work is partly supported by National Science
Foundation  of China under contract Nos. 10075057, 90103020,
19991487, 10225525 and 10147202, the CAS Knowledge Innovation Key
Project KJCX2-SW-N02, and the Deutsche Forschungsgemeinschaft.

\vspace{1cm}
\appendix
{\centerline{\Large \bf{APPENDIX}}}
%\section{APPENDIX}
\vspace{0.5cm}

In this appendix, we give the explicit expressions of $A_{PS, i}(i=1,2,3)$ and $A_{PV,
i}(i=2,3)$  appeared in Eqs.(\ref{appen1}) and (\ref{appen2}),
\begin{eqnarray}
A_{PS,1}&=&(m^2+p \cdot p')~[~(a^2-b^2){(\epsilon \cdot k)}^2+b^2
\epsilon^2 k^2~] \nonumber \\
&~&
-2ab(\epsilon \cdot k)~[~(\epsilon \cdot p)(p' \cdot k)-(\epsilon
\cdot p')
(p \cdot k)~] \nonumber \\
&~&-2b^2~[~\epsilon^2 (p \cdot k)(p' \cdot k)-(\epsilon \cdot k)
(\epsilon \cdot p') (p \cdot k)-(\epsilon \cdot k)(\epsilon \cdot
p)(p' \cdot k)
\nonumber \\
&~&+k^2 (\epsilon \cdot p) (\epsilon \cdot p')~]~,
\end{eqnarray}
\begin{equation}
A_{PS,2}=\frac{1}{m^2}~[~(m^2-p \cdot p')k^2+2(p \cdot k)(p' \cdot k)~]
{\left[ \frac{p'\cdot \epsilon}{2p\cdot k+k^2}
-\frac{p\cdot \epsilon}{2p'\cdot k+k^2} \right]}^2,
\end{equation}
\begin{equation}
A_{PS,3}=4k^2{\left[ \frac{p'\cdot \epsilon}{2p\cdot k+k^2}
-\frac{p\cdot \epsilon}{2p'\cdot k+k^2} \right]}^2~,
\end{equation}
\begin{eqnarray}
A_{PV,2}&=&\frac{1}{m^2}(p \cdot \epsilon-p' \cdot \epsilon)\left(
\frac{p'\cdot \epsilon}{2p\cdot k+k^2}
-\frac{p\cdot \epsilon}{2p'\cdot k+k^2} \right)(p \cdot k +p' \cdot k)
\nonumber \\
&~&+\frac{1}{4m^4}(m^2+p \cdot p'){(p \cdot \epsilon-p' \cdot \epsilon)}^2 ,
\end{eqnarray}
\begin{equation}
A_{PV,3}=\frac{1}{m^2}(p \cdot \epsilon-p' \cdot \epsilon)
\{~ a ~(m^2+p\cdot p')(k \cdot \epsilon)
~+~b~[~ (p' \cdot \epsilon)(p \cdot k)
-(p \cdot \epsilon)(p' \cdot k)~]\}~,
\end{equation}
where we have set for simplicity,
\begin{equation}
a = \frac{1}{2p \cdot k +k^2} - \frac{1}{2p^{\prime} \cdot k + k^2},
\end{equation}
\begin{equation}
b = \frac{1}{2p \cdot k +k^2} + \frac{1}{2p^{\prime} \cdot k + k^2}.
\end{equation}

\begin{small}

\end{small}
\end{document}